# Pressure-induced unconventional charge-density-wave states in kagome metal $AV_3Sb_5$ (A = K, Rb, Cs)

Zhimian Wu[1], Linpeng Nie[1], Ye Yang[3], Kuanglv Sun[1], Huachen Rao[1], Dan Zhao[1], Zhongjun Li[3], Tao Wu[1,2,4,5,*] and Xianhui Chen[1,2,4,5,†]

1. Hefei National Research Center for Physical Sciences at the Microscale, University of Science and Technology of China, Hefei, Anhui 230026, China
2. Department of Physics, University of Science and Technology of China, Hefei, Anhui 230026, China
3. School of Physics, Hefei University of Technology, Hefei 230009, China
4. Collaborative Innovation Center of Advanced Microstructures, Nanjing University, Nanjing 210093, China
5. Hefei National Laboratory, University of Science and Technology of China, Hefei 230088, China

*wutao@ustc.edu.cn
†chenxh@ustc.edu.cn

## Abstract:

**Since the discovery of charge density wave (CDW) and superconductivity, kagome metal $AV_3Sb_5$ (A = K, Rb, Cs) provides a new platform for exploring novel many-body quantum phenomena. In $CsV_3Sb_5$, a stripe-like CDW with commensurate wave vector $q$ = 3/8 was observed under moderate pressures, which leads to a peculiar superconducting double-dome behavior in pressure-dependent phase diagram. Previous density functional theory (DFT) calculations indicate that the pressure-induced stripe-like CDW is beyond conventional phonon softening scenario, suggesting a nontrivial role of electronic correlations. However, an in-depth understanding for the pressure-induced unconventional CDW remains elusive. Here, we performed pressure-dependent $^{51}V$ nuclear magnetic resonance (NMR) measurements on $KV_3Sb_5$ and $RbV_3Sb_5$. Although the superconducting double-dome behavior is absent in pressurized $KV_3Sb_5$ and $RbV_3Sb_5$, a pressure-induced CDW phase, ascribed to a possible incommensurate triple-$Q$ CDW, is identified by NMR spectra in both materials, indicating that the pressure-induced unconventional CDW beyond DFT calculations is a common feature for kagome metal $AV_3Sb_5$. In contrast to the stripe-like CDW, the pressure-induced incommensurate triple-$Q$ CDW does not strongly suppress the superconducting temperature ($T_c$) but coincide with an almost plateau behavior at intermediate pressure regime in the pressure-dependent superconducting phase diagram. Furthermore, by systematically analyzing the Korringa relation between Knight shift and nuclear spin-lattice relaxation rate in $AV_3Sb_5$, van Hove singularities (vHSs) driven electronic fluctuations are revealed as an effective knob for the** **pressure-induced unconventional CDW. Finally, our present findings underscore the pressure-induced unconventional CDW as a novel correlated quantum state in kagome metal $AV_3Sb_5$.**

## I. INTRODUCTION

Vanadium-based kagome metals $AV_3Sb_5$ (A = K, Rb, and Cs) have recently emerged as a versatile platform for exploring novel many-body quantum phenomena within geometrically frustrated lattices [1-4]. In this family, a triple-$Q$ charge-density-wave (CDW) order emerges below $T_{CDW} \approx 78 \sim 102$ K [2-4]. Crucially, this CDW state is unconventional, exhibiting intrinsic symmetry-breaking characteristics such as time-reversal symmetry breaking [5-15] and rotational symmetry breaking [16–21]. Upon further cooling, the electronic system evolves into a superconducting ground state [2–4], which is strongly affected by the preexisting CDW [8,9,12,18,22-35]. Despite extensive research, the microscopic origin of this complex CDW state and its interplay with the superconducting condensate remain subjects of intense debate [5,14,18,27,36-69].

The complexity of the electronic ground state is largely attributed to the unique band structure of the kagome lattice. Recent experiments have identified multiple saddle points near the Fermi level ($E_F$) in these kagome metals [38,39,47-49,70]. The inherent nature of these van Hove singularities (vHSs) induces hopping interference effects, which significantly enhance electronic correlations and drive the formation of diverse electronic orders [36,71]. On one hand, theoretical studies based on correlated electronic models suggest that Fermi surface nesting between vHSs drives the CDW transition [38-42,48,51,60,72,73]. This nesting enhances non-local electronic correlations [74-76], which may subsequently lead to unconventional superconducting pairing [28,32,35,51,74,77]. On the other hand, density functional theory (DFT) calculations highlight the critical role of electron-phonon coupling (EPC), evidenced by imaginary phonon frequencies at the $M$ and $L$ points [40,50,54,73,78-80]. Indeed, a growing body of experimental evidence from spectroscopic studies [50,53,54,57,59,63,81,82] and first-principles calculations [57,64,67,83] has underscored the essential contributions of strong EPC and lattice anharmonicity to CDW formation [57,63,64,67,68,81-85]. Notably, X-ray scattering experiments have unambiguously revealed acoustic phonon softening in $KV_3Sb_5$ [85] and, remarkably, even in $CsV_3Sb_5$, where such behavior had long remained elusive [68].

Regarding superconductivity (SC), while the precise pairing mechanism remains controversial [18,28,32,35,57,58,65,66,86-89], recent studies have increasingly highlighted the significant role of EPC [50,58,64-67,83]. Consequently, a central challenge in kagome superconductors is to elucidate the cooperative effects of both electronic correlations and EPC in shaping the CDW and its interplay with superconductivity. Because these quantum phases are highly sensitive to the specific alkali metal ion [2–4], the $AV_3Sb_5$ family provides an ideal platform for systematic comparative studies. Theoretical calculations indicate that the vHS filling level in $CsV_3Sb_5$ is closest to $E_F$ [84,90,91], suggesting it should host the most prominent electronic correlations [90]. However, reported correlation strengths across the series vary significantly [84,92], likely stemming from sample-dependent variations in $E_F$. This intrinsic diversity underscores the necessity of systematic investigations to unravel the mechanisms governing the electronic states within the $AV_3Sb_5$ family.

Hydrostatic pressure serves as a powerful tuning parameter for investigating the interplay between SC and the CDW. Although pressure uniformly suppresses the primary CDW order across the $AV_3Sb_5$ family [8,22,23], the superconducting behaviors diverge significantly: $CsV_3Sb_5$ exhibits a double-dome superconducting phase diagram [8], whereas $KV_3Sb_5$ and $RbV_3Sb_5$ show a monotonic enhancement of the critical temperature ($T_c$) [22,23]. Recent studies have attributed this anomalous SC evolution in $CsV_3Sb_5$ to its competition with a pressure-induced CDW phase, identified as a stripe-like CDW by nuclear magnetic resonance (NMR) [24] or a long-range 3/8 charge-ordered phase by X-ray diffraction (XRD) [30]. Despite these advances, the pressure-dependent CDW evolution in $KV_3Sb_5$ and $RbV_3Sb_5$ remains largely unexplored. Previous transport measurements indicate that the maximum $T_c$ is achieved at approximately 0.5 GPa in $KV_3Sb_5$ and 1.5 GPa in $RbV_3Sb_5$ [22,23]. Notably, while the CDW transition vanishes above 0.5 GPa in $KV_3Sb_5$ [22], it persists up to 2.4 GPa in $RbV_3Sb_5$ [23]. This discrepancy suggests a complex interplay between the CDW and SC in $RbV_3Sb_5$, raising the possibility that a hidden CDW phase may emerge in the pressure regime between 1.5 GPa and 2.4 GPa.

As a powerful local probe, NMR provides crucial microscopic insights into charge modulations. In this work, we systematically investigated the pressure-induced evolution of the CDW state in $KV_3Sb_5$ and $RbV_3Sb_5$ using $^{51}V$ NMR. Through an analysis of the spectral line shape evolution, we uncovered a pressure-induced incommensurate CDW phase in both compounds. Furthermore, once the incommensurate CDW phase is fully suppressed, the distinct deviations in the Knight shift and spin-lattice relaxation rate ($1/T_1T$) among $AV_3Sb_5$ highlight the pivotal role of vHSs in shaping the pressure-induced unconventional CDW phase and its interplay with superconductivity.

## II. RESULTS

## A. Pressure-dependent superconductivity and CDW state

At ambient pressure, the $^{51}$V NMR spectrum of $AV_3Sb_5$ exhibits seven transition lines with the magnetic field applied along the $c$-axis above $T_{CDW}$. Upon entering the CDW phase, the kagome lattice reconstructs into a staggered tri-hexagonal superlattice [Fig. 1(a) and (b)], generating two inequivalent vanadium sites within the hexagon cluster (V(I)) and the triangle cluster (V(II)). Correspondingly, the $^{51}$V NMR transition lines split into two sets of peaks with equal intensity [Fig. 1(c)] (see Fig. 8 for full spectra). As elucidated by our previous NMR experiments [93], the contribution of the second-order quadrupole effect to the resonance frequency of the central transition lines is negligible. Therefore, this equal splitting is predominantly driven by the magnetic Knight shift ($K = (f - \gamma H)/\gamma H$). In addition to the major splitting, both V(I) and V(II) sites develop a minor splitting with an intensity ratio of 2:1 [Fig. 1(c)], which is attributed to a bulk $C_2$ distortion arising from interlayer stacking [Fig. 1(b)]. For simplicity, we focus primarily on the central transition line to investigate the pressure-dependent spectral evolution.

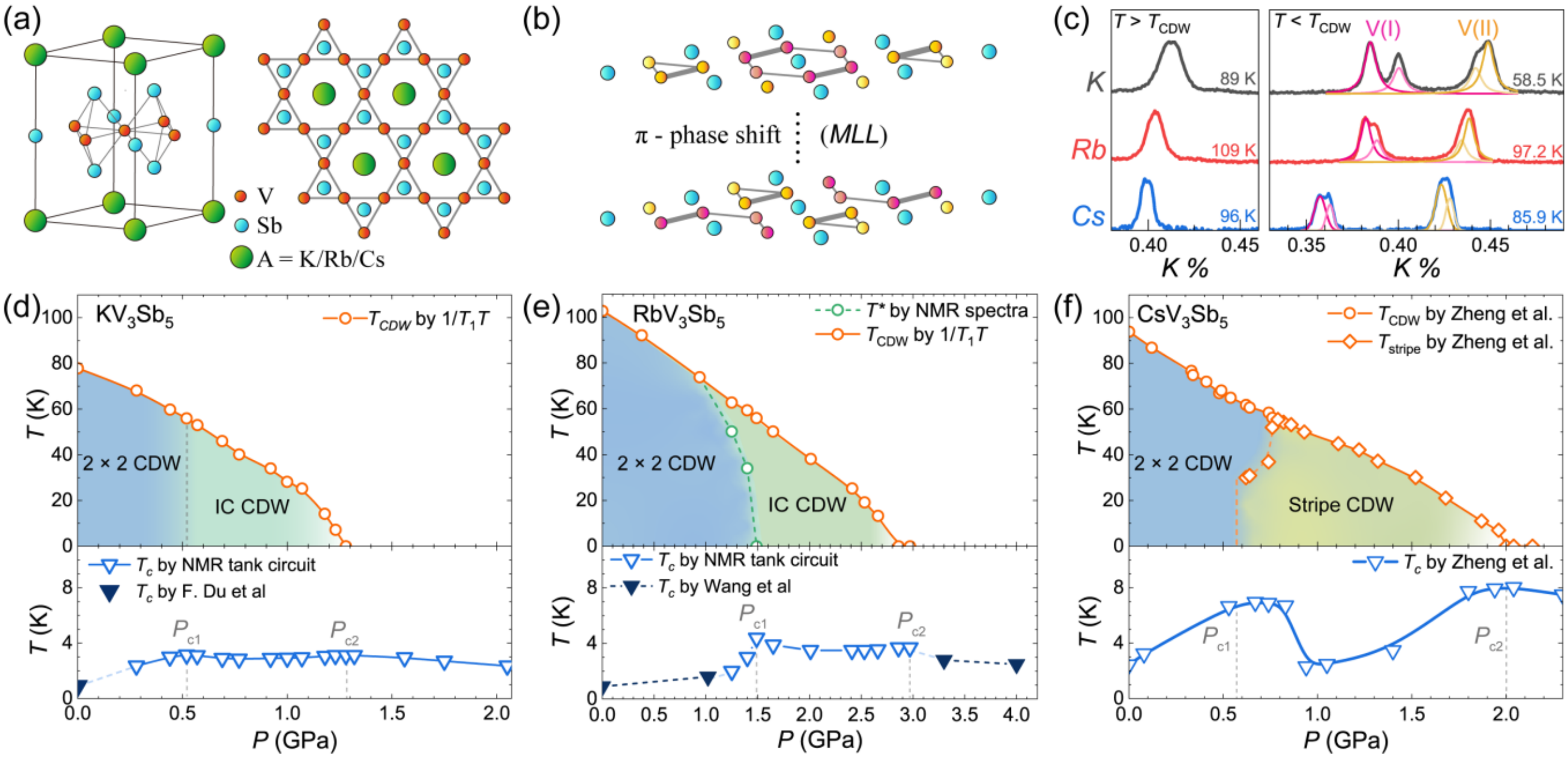

**FIG. 1. Crystal structures and pressure-temperature (*P-T*) phase diagrams of $AV_3Sb_5$ (A = K, Rb, and Cs). (a)** The kagome lattice structure of the $AV_3Sb_5$ family. **(b)** Charge ordering pattern of the vanadium layers in the staggered tri-hexagonal structure. **(c)** $^{51}$V NMR central transition lines of $AV_3Sb_5$ (A = K, Rb, Cs) measured above and below $T_{CDW}$. The pink and orange curves correspond to the vanadium sites in the hexagon cluster (V(I)) and the triangle cluster (V(II)), as depicted in **(b)**. Each spectrum is fitted with four components, where the intensity ratio of the major splitting follows V(I):V(II) = 1:1, and the minor splitting on both V(I) and V(II) follows a 1:2 ratio. P-T phase diagrams for **(d)** $KV_3Sb_5$, **(e)** $RbV_3Sb_5$, and **(f)** $CsV_3Sb_5$, summarizing the evolution of CDW states and superconductivity discussed in the text. The characteristic temperatures are defined in the preceding figures. Data in **(f)** are adopted from our previous work on $CsV_3Sb_5$ [24].

As illustrated in Fig. 1(d)–(f), the $T_c$ is characterized by the temperature-dependent relative change in the resonant frequency ($-\Delta f/f$) of the NMR tank circuit at zero magnetic field (see Fig. 9 for details). With increasing pressure, the $T_c$ of both compounds is gradually enhanced, reaching a maximum of 3.15 K at 0.52 GPa for $KV_3Sb_5$ and 4.38 K at 1.49 GPa for $RbV_3Sb_5$. Notably, above these critical pressures ($P_{c1}$), $T_c$ exhibits only a slight decrease in both compounds. This behavior stands in stark contrast to the rapid suppression of $T_c$ observed in $CsV_3Sb_5$ [Fig. 1(f)] [24]. These results are consistent with previous reports [22,23], confirming a single-dome evolution of pressure-dependent superconductivity in $KV_3Sb_5$ and $RbV_3Sb_5$. In $CsV_3Sb_5$, the emergence of a stripe-like CDW shapes a double-dome superconducting phase diagram. Therefore, elucidating the microscopic nature of the CDW and its evolution under pressure is essential for understanding the superconducting landscape in $KV_3Sb_5$ and $RbV_3Sb_5$.

Previous transport measurements indicate that while the CDW transition persists up to 2.4 GPa in $RbV_3Sb_5$, it vanishes above 0.5 GPa in $KV_3Sb_5$ [22,23]. As shown in Fig. 2(a) and (b), the characteristic two-peak structure of the $^{51}$V NMR spectra is preserved in both $KV_3Sb_5$ and $RbV_3Sb_5$ below $P_{c1}$. However, distinct anomalies emerge at $P_{c1}$: in $KV_3Sb_5$, both peaks broaden toward the central frequency, whereas in $RbV_3Sb_5$, a "third peak" abruptly appears between the original two peaks. Above $P_{c1}$, the spectra in both compounds evolve continuously, forming broad "bell-like" line shapes. Eventually, the spectra restore the single-peak structure characteristic of a perfect kagome lattice above $P_{c2}$ = 1.28 GPa for $KV_3Sb_5$ and $P_{c2}$ = 2.86 GPa for $RbV_3Sb_5$, signaling the complete suppression of the CDW order. These phenomena bear a strong resemblance to our previous work on pressurized $CsV_3Sb_5$ [Fig. 2(c)] [24], where a new spectral line shape replaced the two-peak structure above $P_{c1}$, signaling the emergence of a new CDW phase (identified as a stripe-like order with unidirectional 8/3 $a_0$ modulation) [Fig. 1(f)]. Furthermore, these spectral anomalies align with recent studies suggesting the emergence of distinct pressure-induced phases in $KV_3Sb_5$ and $RbV_3Sb_5$ [94,95]. Consequently, the remarkable changes in spectral line shape observed near $P_{c1}$ imply that the maximum $T_c$ coincides with the boundary between two CDW phases, suggesting a complex interplay between the emergent CDW and superconductivity.

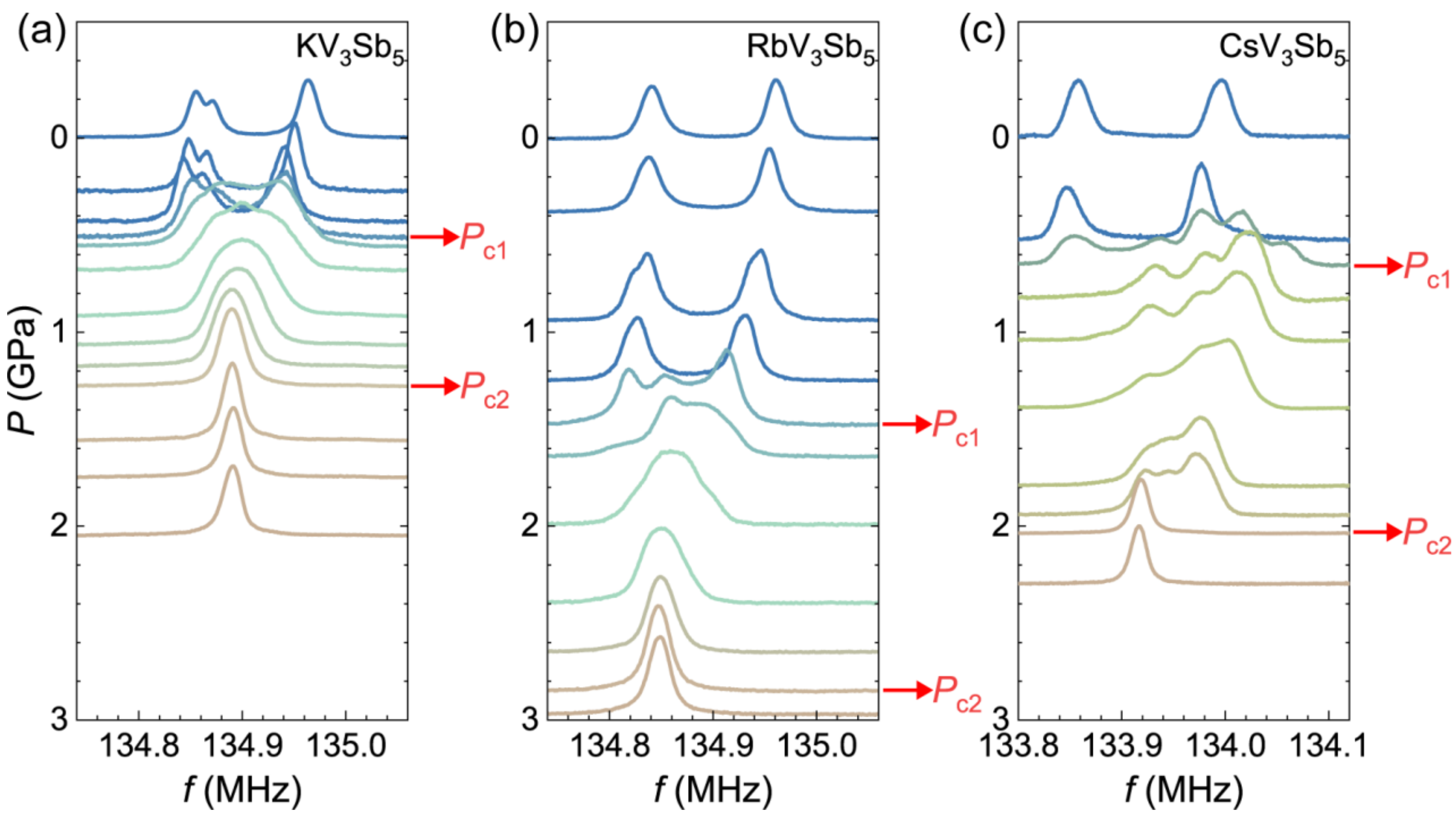

**FIG. 2. Pressure-dependent evolution of the CDW states in $AV_3Sb_5$ (A = K, Rb, and Cs). (a)** $KV_3Sb_5$, **(b)** $RbV_3Sb_5$ and **(c)** $CsV_3Sb_5$. All $^{51}$V central transition lines are measured at 2 K with a magnetic field of 12 T applied along the crystallographic $c$ - axis. The red arrows indicate the critical pressures for the CDW transitions and suppression. Data in **(c)** are adopted from our previous work on $CsV_3Sb_5$ [24].

## B. Temperature-dependent evolution of CDW state under different pressures

Figure 3 presents the temperature-dependent NMR spectra for $KV_3Sb_5$ and $RbV_3Sb_5$ under various pressures. Below $P_{c1}$, the spectra consistently split into two peaks at low temperatures [Fig. 3(b) and (g)], corresponding to the preservation of the triple-$Q$ CDW state. Notably, a two-step CDW transition is observed near $P_{c1}$ in $RbV_3Sb_5$. In Fig. 3(g), the spectra at 1.4 GPa first broaden below ~60 K, forming a distinct line shape. Upon further cooling, the two peaks characteristic of the triple-$Q$ CDW emerge below 40 K, eventually restoring the standard two-peak structure at 30 K. At $P_{c1}$, while the splitting persists in $KV_3Sb_5$, both peaks broaden asymmetrically toward the central frequency [Fig. 3(c)]. In $RbV_3Sb_5$, the new spectral profile coexists with the characteristic two-peak structure down to 2 K [Fig. 3(h)], indicating the coexistence of the triple-$Q$ CDW and the new CDW phase.

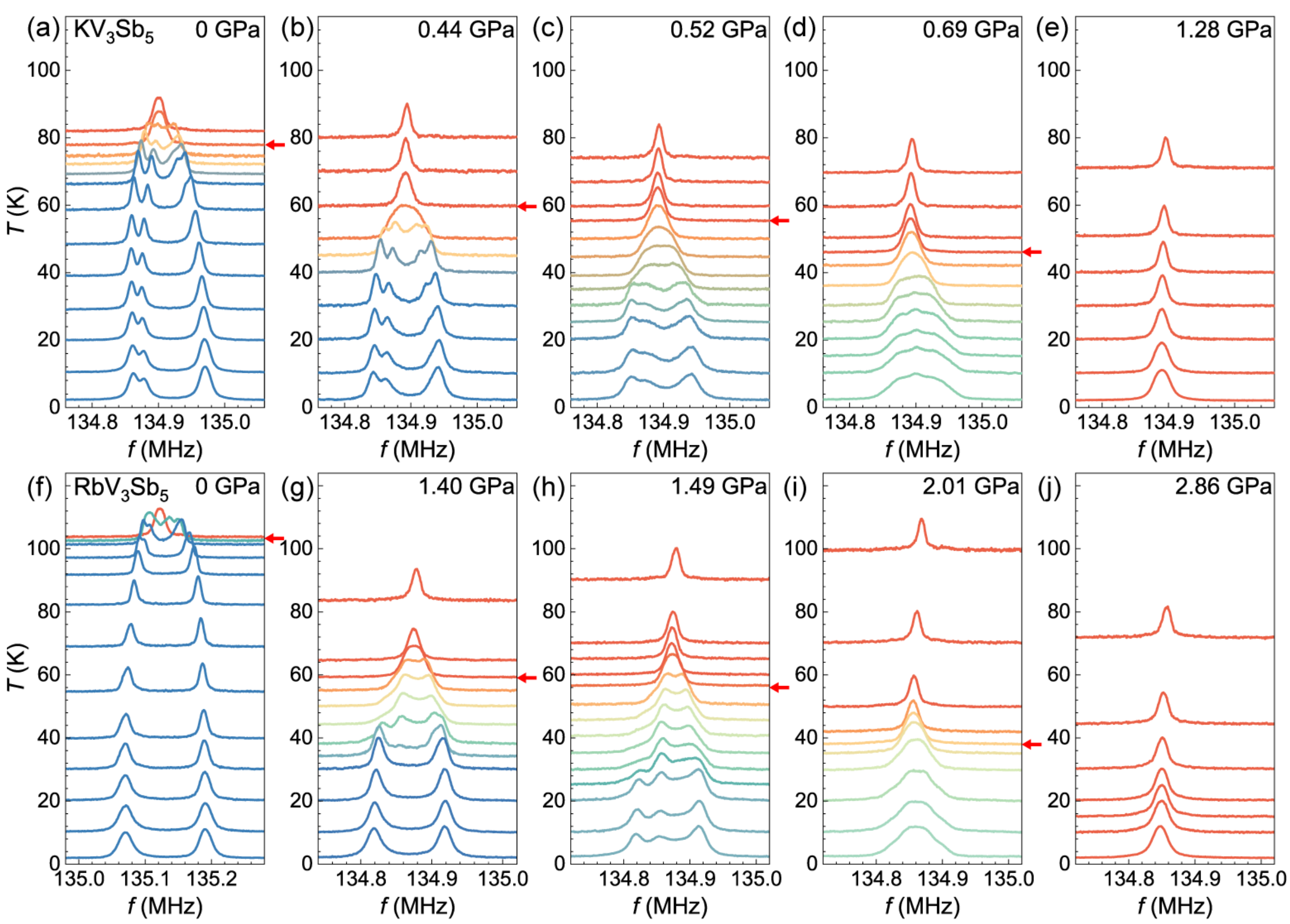

**FIG. 3. Pressure-dependent multiple CDW transitions revealed by $^{51}$V NMR spectra. (a)–(e)** Temperature-dependent $^{51}$V NMR spectra of $KV_3Sb_5$ measured under pressures of 0, 0.44, 0.52, 0.69, and 1.28 GPa. **(f)–(j)** Temperature-dependent V NMR spectra of $RbV_3Sb_5$ measured under pressures of 0, 1.4, 1.49, 2.01, and 2.86 GPa. The red, blue, and green lines correspond to the kagome phase, the triple-$Q$ CDW phase, and the incommensurate CDW phases, respectively. The red arrows indicate the CDW transition temperatures, as determined by $1/T_1T$ (see details in Fig. 4 and Supplementary Section III [96]).

As pressure increases further above $P_{c1}$, the characteristic two-peak structure eventually disappears. The typical spectral line shapes for the second CDW phase in $KV_3Sb_5$ and $RbV_3Sb_5$ are shown in Fig. 3(d) and (i). Here, the single peak gradually broadens upon cooling, establishing a "bell-like" line shape at 2 K, which confirms the establishment of a new emergent CDW phase. At higher pressures, the CDW transition becomes difficult to discern due to the diminished spectral linewidth at low temperatures (see Supplementary Section II[96]). To further probe the electronic state, we measured the temperature-dependent nuclear spin-lattice relaxation rate $(1/T_1)$ of the $^{51}$V nuclei [Fig. 4]. Generally, the relaxation rate is described by $1/T_1T = (1/T_1T)^{QP} + (1/T_1T)^{SF}$, where the first term arises from quasiparticle contributions and the second from spin fluctuations. As shown in Fig. 4(a) and (f), at ambient pressure, $1/T_1T$ decreases rapidly below the CDW transition temperature due to the opening of a momentum-dependent CDW gap, as observed in previous ARPES studies [19,31,38]. Consistently, we define the onset temperature of this drop in $1/T_1T$ as $T_{CDW}$ [Fig. 4], which gradually decreases with increasing pressure in both compounds (see Supplementary Section III for full characterization [96]). As shown in Fig. 4(e) and (j), the gap behavior vanishes completely at $P_{c2}$ = 1.28 GPa for $KV_3Sb_5$ and $P_{c2}$ = 2.86 GPa for $RbV_3Sb_5$, confirming the full suppression of the CDW order.

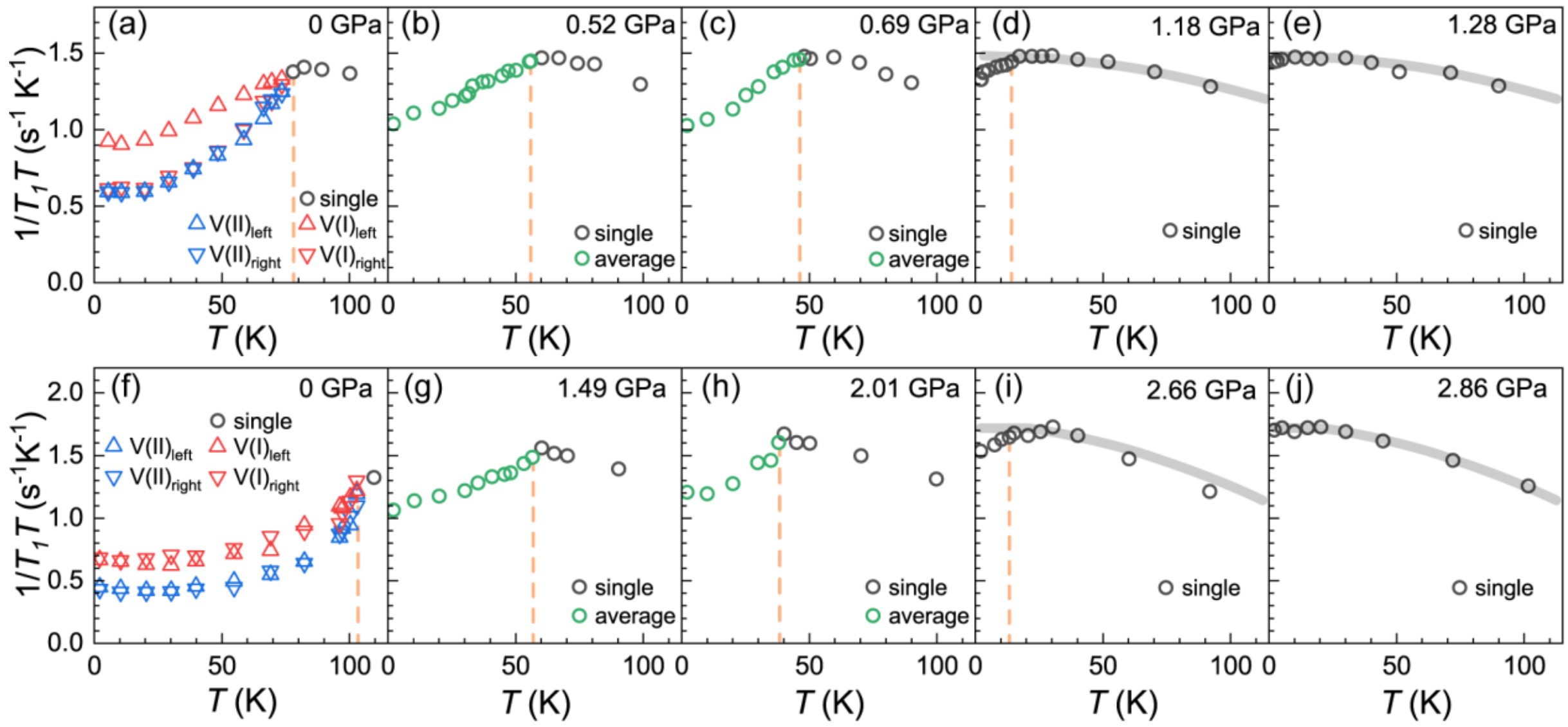

**FIG. 4. Evidence for multiple CDW transitions from nuclear spin-lattice relaxation measurements. (a)–(d)** Temperature-dependent $1/T_1T$ behaviors of $^{51}$V NMR in $KV_3Sb_5$ under various pressures. **(e)–(h)** Temperature-dependent $1/T_1T$ behaviors of $^{51}$V NMR in $RbV_3Sb_5$ under various pressures. The orange dashed lines delineate $T_{CDW}$, defined by the onset of the drop in $1/T_1T$. Near $P_{c2}$, where the drop in $1/T_1T$ becomes obscure, $T_{CDW}$ is determined by the deviation from the high-temperature behavior, as indicated by the grey dashed lines in **(d)**, **(e)**, **(i)**, and **(j)**.

## C. Identification of pressure-induced incommensurate CDW states

Characterizing the second CDW phases emerging between $P_{c1}$ and $P_{c2}$ remains a significant challenge. In contrast to the five discrete peaks observed in the stripe-like CDW state of $CsV_3Sb_5$ [24], the spectral profiles for $KV_3Sb_5$ and $RbV_3Sb_5$ in this regime manifest as broad, featureless "bell-like" shapes. Generally, in a commensurate CDW state, each characteristic NMR peak corresponds to a specific inequivalent nuclear site; consequently, the number of peaks and their intensity ratios provide a reliable basis for identifying discrete CDW models via spectral fitting. However, in an incommensurate CDW state, the loss of translational periodicity generates a continuum of inequivalent nuclear sites [97]. This absence of discrete sites likely accounts for the lack of distinct peaks and the pronounced spectral broadening observed in Figs. 2(a) and (b) between $P_{c1}$ and $P_{c2}$.

Nevertheless, despite the absence of sharp discrete peaks, a closer inspection reveals subtle yet highly reproducible features within these broad profiles below $T_{CDW}$. For $KV_3Sb_5$, the spectra at 0.69 GPa [Fig. 3(d)] and 0.77 GPa [Fig. S2(g)] consistently exhibit a small central protrusion. For $RbV_3Sb_5$, the spectra below $T_{CDW}$ at 2.01 GPa [Fig. 3(i)] display a central plateau flanked by two distinct inflection points. Remarkably, these line-shape characteristics bear a striking resemblance to the NMR profiles theoretically predicted for incommensurate systems governed by triple-$Q$ modulation [97], as illustrated in Figs. 5(a) and (d). This close correspondence strongly suggests an incommensurate triple-$Q$ CDW origin and motivates a systematic numerical simulation using various multiple-$Q$ models to assess the consistency of the different models with the experimental data.

To investigate this further, we applied several incommensurate CDW models to simulate the spectral line shapes of $KV_3Sb_5$ and $RbV_3Sb_5$ (see Appendix A for details). The single-Q and double-Q models considered here do not reproduce the key spectral features, yielding either a deep central gap or a disproportionate central weight (see Fig. 11 for details). Instead, triple-$Q$ incommensurate CDW modulations incorporating distinct in-plane anisotropies provide the most consistent interpretation of the experimental data. As shown in Fig. 5(c), the simulated profile excellently reproduces the experimental spectrum of $KV_3Sb_5$ at 0.69 GPa. The optimal fit was achieved with a CDW amplitude ratio of 1:1:2, suggesting that while the macroscopic triple-$Q$ symmetry is preserved, the local charge distribution exhibits pronounced anisotropy. Similarly, for $RbV_3Sb_5$ at 2.01 GPa, the experimental spectrum aligns with a simulated profile defined by an amplitude ratio of 2:1:1 [Fig. 5(f)].

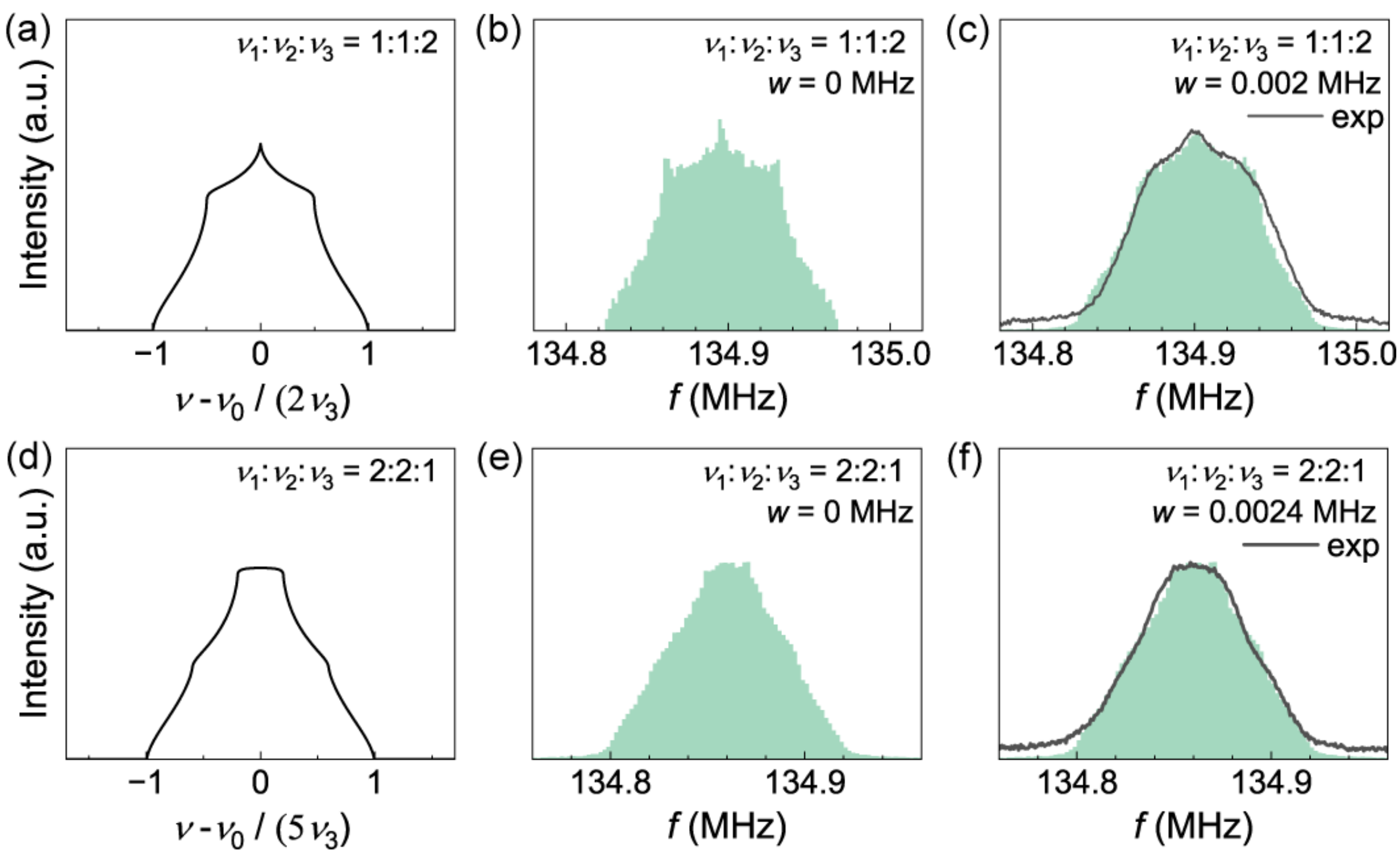

**FIG. 5. Simulation of the pressure-induced incommensurate CDW states. (a)** Theoretical NMR line shape for a triple-$Q$ modulation with relative amplitudes $2\nu_1 = 2\nu_2 = \nu_3$. **(b)** The standard simulated NMR line shape as the same modulation in **(a)** without broadening. **(c)** Comparison between the simulated NMR line shape for triple-$Q$ plane wave modulations (with parameters $\nu_0 = 134.901\text{MHz}$ and $2\nu_1 = 2\nu_2 = \nu_3 = 0.036\text{MHz}$) and the experimental spectra of $KV_3Sb_5$ at 2 K and 0.69 GPa. **(d)** Theoretical NMR line shape for a triple-$Q$ modulation with relative amplitudes $\nu_1 = \nu_2 = 2\nu_3$. **(e)** The standard simulated NMR line shape as the same modulation in **(d)** without broadening. **(f)** Comparison between the simulated NMR line shape for triple-$Q$ plane wave modulations (with parameters $\nu_0 = 134.86\text{MHz}$ and $\nu_1 = \nu_2 = 2\nu_3 = 0.024\text{MHz}$) and the experimental spectra of $RbV_3Sb_5$ at 2 K and 2.01 GPa. The theoretical NMR line shapes in **(a)** and **(d)** are cited from ref [97].

Consequently, the bell-like line shape is consistent with, and supports, the proposed interpretation of an anisotropic triple-Q incommensurate CDW. In summary, our comprehensive analysis of the $^{51}V$ NMR spectra identifies this anisotropic incommensurate triple-$Q$ CDW as the most plausible candidate for the second CDW phase in both $KV_3Sb_5$ and $RbV_3Sb_5$. While NMR serves as a sensitive local probe for these charge modulations, determining the precise modulation period and detailed long-range crystallographic structure of these incommensurate phases will require future scattering measurements, such as high-pressure X-ray diffraction (XRD).

### D. Pressure-dependent phase diagram of the $AV_3Sb_5$ family

Figures 1(a)–(c) summarize the pressure-dependent ($P$) phase diagrams of the $AV_3Sb_5$ family. On one hand, clear commonalities exist across all three compounds, most notably the direct competition between SC and the triple-$Q$ CDW below $P_{c1}$, followed by the emergence of a second CDW phase between $P_{c1}$ and $P_{c2}$. On the other hand, striking divergences emerge in the characterization of these second CDW phases and their respective interplay with superconductivity. This sharp contrast in phase evolution raises intriguing questions regarding the underlying physics governing these pressure-induced CDW states.

Previous theoretical analyses suggest that unstable phonon modes at the $M$ and $L$ points drive the triple-$Q$ CDW order at ambient pressure [40,50,54,73,78-80]. Within this framework, a commensurate-to-incommensurate transition is expected if the minimum of the imaginary phonon frequency shifts away from these high-symmetry points. To test this, we performed first-principles calculations of the pressure-dependent phonon spectra. As shown in Fig. 12, the critical pressure at which the imaginary frequency disappears aligns closely with the suppression of the CDW in $(K,Rb)V_3Sb_5$. However, the commensurate-to-incommensurate transition is notably absent in our calculations. Specifically, the minimum of the imaginary phonon frequency remains fixed at the $M$ and $L$ points until the instability vanishes, failing to reproduce the experimental observations. Furthermore, detailed characteristics of the incommensurate phase, such as the varying relative amplitudes of CDW modulation across different wave vectors, deviate significantly from standard weak-coupling scenarios. These discrepancies underscore that a pure EPC driven mechanism is insufficient to explain the pressure-induced incommensurate CDW, identifying it instead as an unconventional correlated state beyond the scope of standard DFT descriptions.

The limitations of the purely phonon-driven scenario discussed above lead us to consider the critical role of electronic correlations. Interestingly, the nearly plateau-like behavior of $T_c$ observed within the incommensurate CDW phases of $KV_3Sb_5$ and $RbV_3Sb_5$ parallels observations in transition metal dichalcogenides (TMDs), where a superconducting dome coincides with a commensurate-to-incommensurate transition [98,99]. Recent theoretical work on monolayer $1T$-$TiSe_2$ has demonstrated that the suppression and incommensuration of the CDW cannot be accurately captured by phonon-driven scenarios alone but crucially require the inclusion of on-site electronic correlations [100]. Specifically, these correlations dominate the phonon-driven instability by renormalizing the momentum-dependent electron-phonon coupling, allowing for an accurate theoretical reproduction of the experimental phase diagram, including both commensurate and incommensurate regions. This suggests that the incommensurate CDW is an electronic state inherently governed by correlation effects, which may play a fundamental role in supporting the superconducting phase.

Therefore, we propose that in (K,Rb)$V_3Sb_5$, non-local electronic correlations promoted by vHSs cooperate with the EPC to drive CDW evolution under pressure. Additionally, in $CsV_3Sb_5$, the triple-$Q$ CDW evolves into a unidirectional stripe-like modulation; this transformation exceeds simple changes in band structure or phonon dispersion and likely reflects a regime dominated by more prominent electronic correlation effects [24]. Crucially, the interplay with superconductivity also shifts from the coexistence observed in $KV_3Sb_5$ and $RbV_3Sb_5$ to a regime of strong competition in $CsV_3Sb_5$. Considering that the electronic structures of $KV_3Sb_5$ and $RbV_3Sb_5$ are more similar to each other than to $CsV_3Sb_5$ [84,91], we suggest that the divergent CDW evolution paths across the $AV_3Sb_5$ family are mediated by the vHS alignment, which acts as an effective knob to tune the correlation strength.

### E. The role of vHS on electronic correlations

To explore the evolution of electronic correlations across the $AV_3Sb_5$ family, we systematically compared the temperature-dependent Knight shift [Fig. 6(a)] and $1/T_1T$ [Fig. 6(b)] behaviors at $P_{c2}$ where the CDW phase is completely suppressed. Generally, the total Knight shift consists of a temperature-independent orbital term ($K_{\rm orb}$) and a temperature-dependent spin term ($K_{\rm spin}$). Since $K_{orb}$ remains constant at a fixed pressure without structural or magnetic phase transitions, the observed temperature variation of the Knight shift is strictly governed by $K_{spin}$. This spin component is proportional to the density of states (DOS) at the Fermi level, $N(E_F)$: $K_{spin} = A_{hf}\mu_B N(E_F)$ (where $A_{hf}$ is the hyperfine form factor and $\mu_B$ is the Bohr magneton). In a weakly interacting Fermi liquid where additional spin fluctuations are negligible, $1/T_1T$ reflects $N(E_F)$ and follows the standard Korringa relation, $1/T_1T \propto K_{spin}^2 \propto N(E_F)^2$. Conversely, when quasiparticle interactions are significant, electronic instabilities can induce enhanced fluctuations that strongly renormalize $1/T_1T$.

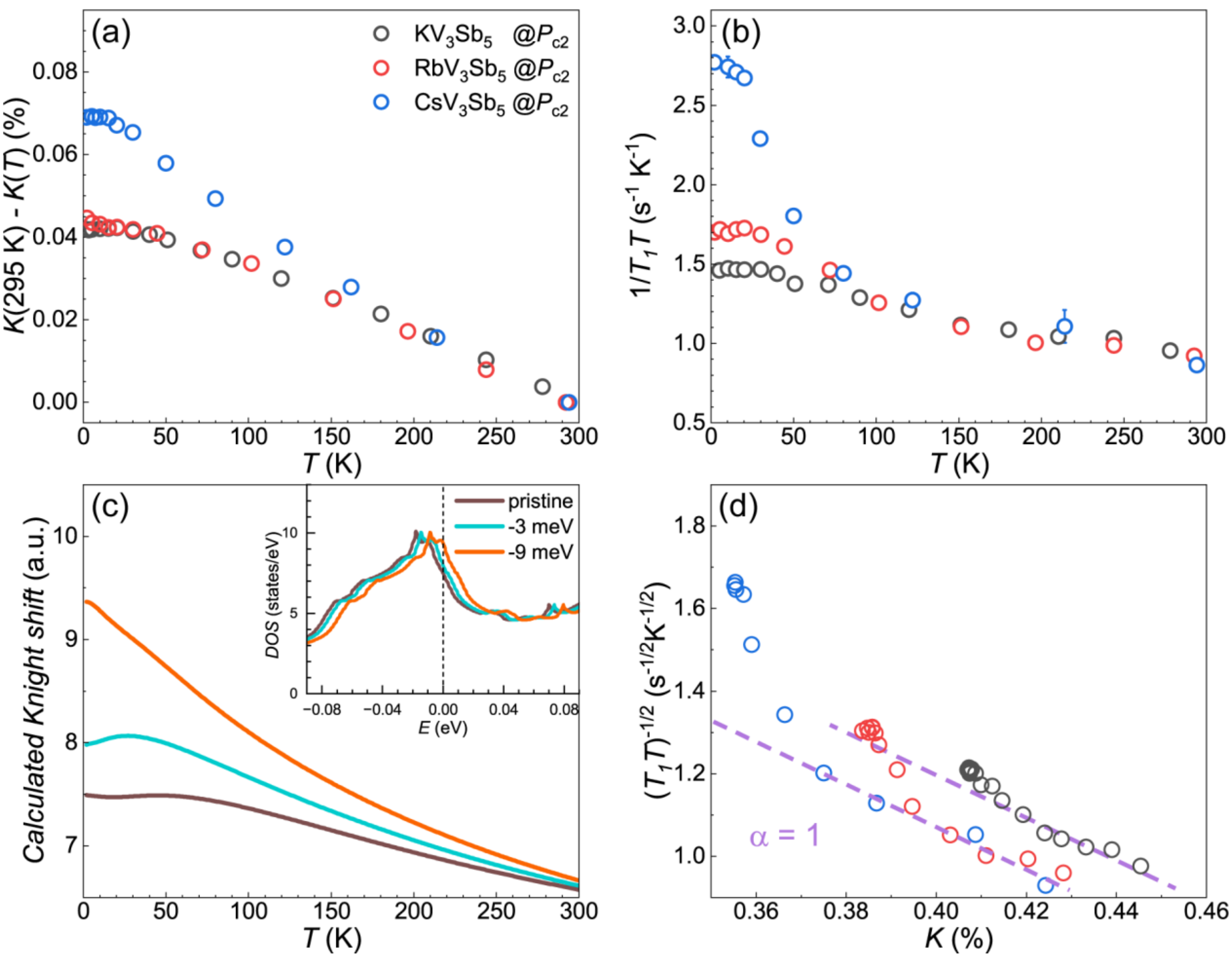

**FIG. 6. Evidence for the role of vHSs across the $AV_3Sb_5$ family, derived from Knight shift and $1/T_1T$ measurements at $P_{c2}$. (a)** Temperature-dependent $^{51}$V NMR Knight shifts in $AV_3Sb_5$ (A = K, Rb and Cs) at $P_{c2}$. The shifts for different compounds are normalized by subtracting the value at approximately 295 K. **(b)** Temperature-dependent $1/T_1T$ of $^{51}$V NMR in $AV_3Sb_5$ (A = K, Rb and Cs) at $P_{c2}$. **(c)** Calculated temperature-dependent Knight shift, proportional to $\frac{1}{k_BT}\int(1-f(E))f(E)D(E)dE$, for different alignments of the van Hove singularities (vHSs) relative to $E_F$. Here, $k_B$ is the Boltzmann constant, $f(E)$ is the Fermi–Dirac distribution function, and $D(E)$ is the density of states (DOS). **(d)** Korringa relation between $1/T_1T$ and the square of the Knight shift ($K_{spin}^2$) in $AV_3Sb_5$ (A = K, Rb and Cs) at $P_{c2}$. The relation is generally expressed as $T_1TK_{spin}^2 = (\hbar\gamma_e^2)/(4\pi k_B\gamma_n^2\alpha)$, where $\gamma_e$ and $\gamma_n$ are the electron and nuclear gyromagnetic ratios, respectively, and $\alpha$ is the Korringa enhancement factor. The violet dashed line indicates the standard Korringa relation with $\alpha = 1$, corresponding to the weak coupling limit.

As shown in Fig. 6(b), the temperature-dependent $1/T_1T$ behaviors in $KV_3Sb_5$ and $RbV_3Sb_5$ differ markedly from that in $CsV_3Sb_5$, which displays a pronounced Curie–Weiss–like upturn upon cooling. According to our previous work, this Curie–Weiss–like behavior in $CsV_3Sb_5$ is pressure-independent and originates from correlation-enhanced electronic fluctuations [24]. This clear discrepancy indicates that electronic correlation effects are relatively weaker in $KV_3Sb_5$ and $RbV_3Sb_5$ than in $CsV_3Sb_5$.

Similarly, the Knight shift in $CsV_3Sb_5$ deviates remarkably from those in $KV_3Sb_5$ and $RbV_3Sb_5$, exhibiting a strong temperature dependence that points to an intrinsically varying $N(E_F)$ [Fig. 6(a)]. We attribute this deviation to the proximity of the vHSs to the Fermi level ($E_F$) in $CsV_3Sb_5$. By tuning the vHS alignment relative to $E_F$ based on previous DFT calculations [24], we successfully reproduced the experimental Knight shift behaviors [Fig. 6(c)] (see Appendix A for details). Prior DFT calculations suggest that while the Fermiology of the $AV_3Sb_5$ family shares similarities near the $M$ point at ambient pressure, the high-order vHS—which exhibits flat dispersion and generates a power-law divergent DOS—is positioned significantly closer to $E_F$ in $CsV_3Sb_5$ than in (K,Rb)$V_3Sb_5$ [70,84,91]. Although recent ambient-pressure angle-resolved photoemission spectroscopy (ARPES) results have struggled to resolve the fine details of vHSs alignment across the family, our NMR measurements provide a unique microscopic probe of the electronic structure in the high-pressure regime where the CDW phase is fully suppressed.

Crucially, our results highlight that the enhancement of electronic correlations in $CsV_3Sb_5$ transcends a simple DOS effect. While the proximity of the vHSs typically increases $N(E_F)$—which can theoretically enhance EPC—it further triggers a synergistic cooperation between the EPC and intrinsic electronic correlations that goes beyond a weak-coupling description. This conclusion is further corroborated by the Korringa plots in Fig. 6(d). While $KV_3Sb_5$ closely follows the standard Korringa behavior and $RbV_3Sb_5$ exhibits a moderate deviation below about 101 K, $CsV_3Sb_5$ shows the largest departure from linearity below about 80 K. This anomalous scaling serves as a clear experimental signature of enhanced electronic fluctuations and the breakdown of the simple quasiparticle picture. Therefore, these results reveal a progressive migration of the vHSs toward $E_F$ as one moves from (K,Rb)$V_3Sb_5$ to $CsV_3Sb_5$, establishing vHSs tuning as the decisive effective knob for scaling the electronic correlation strength across the $AV_3Sb_5$ family.

## III. DISCUSSION

Our discovery of emergent high-pressure CDW phases uncovers a novel regime that defies a simple weak-coupling paradigm, emphasizing the pivotal synergistic interplay between EPC and electronic correlations. Although on-site correlations in $AV_3Sb_5$ systems remain relatively modest, theoretical works [36,41,71,75,76,101-103] have implicated non-local electronic correlations, facilitated by vHSs, as critical stabilizers of charge order. Notably, the present NMR work reveals an intermediate, anisotropic incommensurate CDW phase preceding the complete suppression of CDW order in (K,Rb)$V_3Sb_5$. This phase exhibits distinct modulation amplitudes across the three wave vectors, approximating ratios of 2:1:1 in $KV_3Sb_5$ and 2:2:1 in $RbV_3Sb_5$. These ratios signify an emergent in-plane anisotropy—dissociated from interlayer stacking-induced anisotropy observed at ambient pressure—that prefigures the unidirectional stripe-like modulation inherent to the $AV_3Sb_5$ family. While non-local correlations [24,33] can potentiate this stripe-like tendency, temperature-dependent Knight shift measurements indicate insufficient proximity of the vHSs to the $E_F$ in these compounds. Consequently, electronic correlations lack the requisite strength to rigidify unidirectional stripe distortions, culminating instead in an incommensurate CDW phase. Furthermore, while standard DFT calculations accurately predict the critical pressure for imaginary phonon frequency disappearance aligning with CDW suppression in (K,Rb)$V_3Sb_5$, the observed incommensurability phenomenon eludes explanation by pure EPC models. Therefore, we posit that CDW evolution in (K,Rb)$V_3Sb_5$ is sculpted by cooperative EPC and finite electronic correlations, collectively driving the system away from commensurability.

In stark contrast, the CDW evolution in $CsV_3Sb_5$ deviates fundamentally from phonon-centric scenarios. While standard DFT calculations [Fig. 12] project a CDW instability persisting up to 4 GPa in $CsV_3Sb_5$, experiments establish its full suppression near 2 GPa. This striking disparity implicates a correlation-dominated regime. Notably, $CsV_3Sb_5$ exhibits the greatest departure from the Korringa relation, indicating that its vHSs' closer proximity to $E_F$ triggers intrinsic correlations surpassing mere DOS enhancement. These intensified correlations effectually suppress the initial triple-$Q$ CDW order while stabilizing a unidirectional stripe-like CDW state—a correlated electronic phase emergent from synergistic EPC and vHSs-tuned correlations. It should be noted that recent first-principles calculations also reveal significant lattice anharmonicity in $CsV_3Sb_5$ [64,67], which might be also related to the vHSs and important for the pressure-induced unconventional CDW phases.

Furthermore, vHS-tuned correlations mediate contrasting CDW-superconductivity interplays across the family. In (K,Rb)$V_3Sb_5$, the transition to an incommensurate phase often coincides with the emergence of a superconducting plateau. This coincidence may simply suggest that the pressure-induced CDW and superconductivity in these compounds are only weakly coupled or largely independent. Alternatively, as a possible scenario, analogous phenomena might occur here as in $TiSe_2$ [98,99], where electronic correlations renormalize momentum-dependent EPC, stabilizing incommensurate domain walls compatible with Cooper pairing [100]. Conversely, in $CsV_3Sb_5$, stripe order emergence precipitates a sharp $T_c$ suppression. Recent investigations of emergent superconducting fluctuations [26] report enhanced pairing gaps under pressure in $CsV_3Sb_5$, yet stripe order violently competes with superconducting coherence—echoing cuprate stripe phases, where phase fluctuations robustly suppress macroscopic superconductivity [104]. Moreover, ARPES studies [70] have unveiled a common normal-state electronic structure in $AV_3Sb_5$ deviating from standard DFT predictions, identifying multiple sublattice-pure vHSs near $E_F$, facilitated by significant V-$d$/Sb-$p$ hybridization. Theoretically, this configuration enhances charge and spin bond fluctuations through Fermi surface nesting, favoring bond order or unconventional superconductivity [70], which might be also important for the formation of stripe order in pressurized $CsV_3Sb_5$.

Finally, we propose that the positioning of vHSs relative to $E_F$ functions as a pivotal tuning parameter, governing electronic correlation strength and dictating the selection of dominant ground states—whether loop currents [43,76,101,105-107], chiral excitonic states [108], or correlated stripe phases [24,33]—across the $AV_3Sb_5$ family. This framework unifies disparate phase diagrams in kagome superconductors, highlighting the central role of vHSs in sculpting intertwined electronic orders.

*Acknowledgments* — This work is supported by the National Natural Science Foundation of China (Grant No. 12325403, 12034004, 12161160316, 12488201), the National Key R&D Program of the MOST of China (Grant No. 2022YFA1602601), the Chinese Academy of Sciences under contract No. JZHKYPT-2021-08, the CAS Project for Young Scientists in Basic Research (Grant No. YBR-048), the Innovation Program for Quantum Science and Technology (Grant No. 2021ZD0302800).

# APPENDIX A: MATERIALS AND METHODS

## 1. Samples growth

High-quality single crystals of $KV_3Sb_5$ and $RbV_3Sb_5$ were synthesized via self-flux growth based on previous reports [8]. The high quality of the samples has been verified by the resistivity data in Fig. 7, which were collected on a Quantum Design Physical Properties Measurement System (PPMS). The NMR measurement at 0 GPa and 2.02 GPa is carried on the high-quality single crystal $CsV_3Sb_5$ provided by Dong Chen. All measurements were conducted on the same single crystals to maintain consistency.

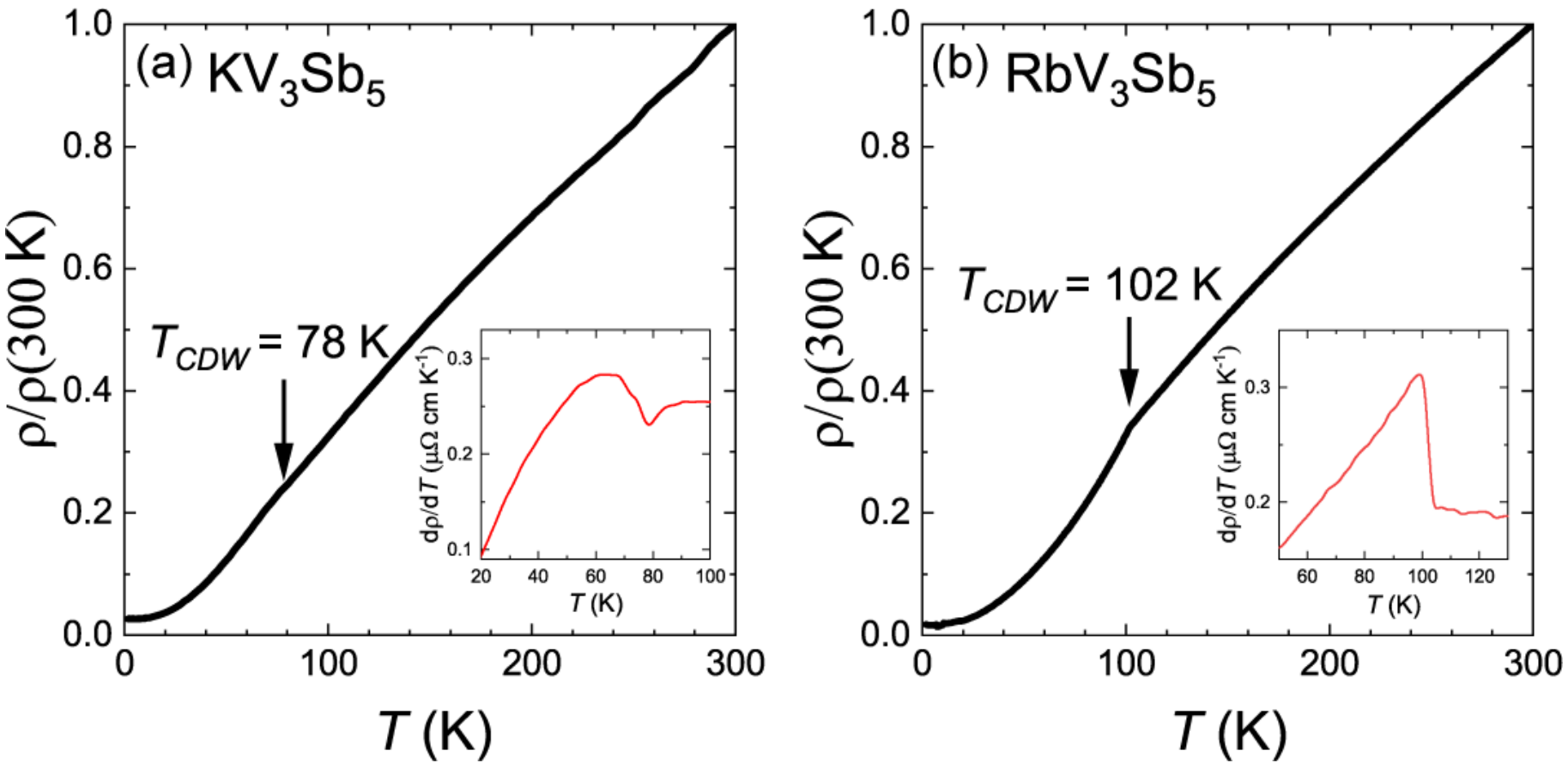

**FIG. 7 *T* - dependent resistivity of $KV_3Sb_5$ and $RbV_3Sb_5$.** The inset plots are the temperature derivative of the in-plane resistivity showing the CDW transitions.

## 2. NMR measurements

A commercial NMR spectrometer from Thamway Co. Ltd. was used for the NMR and NQR measurements. An NMR-quality magnet from Oxford Instruments offers a uniform magnetic field of up to 12 T. The external magnetic field applied on the sample is calibrated by $^{63}$Cu NMR of the NMR coil. All NMR measurements were performed under an external magnetic field at 12 T along the c axis. The nuclear spin-lattice relaxation times ($T_1$) of $^{51}$V NMR and $^{121}$Sb NQR were measured via the inverse recovery method. The fitting function of the recovery of the nuclear magnetization M(t) for $^{51}$V NMR is $1-\frac{M(t)}{M(\infty)} = I_0\left(\frac{1}{84}\times\exp\left(-\left(\frac{t}{T_1}\right)^\beta\right)+\frac{3}{44}\times\exp\left(-\left(\frac{6t}{T_1}\right)^\beta\right)+\frac{75}{364}\times\exp\left(-\left(\frac{15t}{T_1}\right)^\beta\right)+\frac{1225}{1716}\times\exp\left(-\left(\frac{28t}{T_1}\right)^\beta\right)\right)$ [109], whereas that for $^{121}$Sb NQR on the $\left(\pm\frac{5}{2}\leftrightarrow\pm\frac{3}{2}\right)$ transition is $1-\frac{M(t)}{M(\infty)} = I_0\left(0.571\times\exp\left(-\left(\frac{10t}{T_1}\right)^\beta\right)+0.428\times\exp\left(-\left(\frac{3t}{T_1}\right)^\beta\right)\right)$ [110].

## 3. Calibration of the hydrostatic pressure

The hydrostatic pressure is realized by self-clamped BeCu piston-cylinder cell. The pressure transmitting medium is Daphne 7373 oil. The pressure was calibrated through the $^{63}$Cu NQR frequency of $Cu_2O$ powder in Fig. S1 [96]. To ensure accurate pressure determination, we employed a dual-coil configuration inside the sample chamber. The pressure manometer ($Cu_2O$ NQR coil) was positioned just below the sample coil, with a vertical spatial separation of less than 2 mm. This minimal distance eliminates any significant spatial pressure gradient between the sample and the pressure sensor, which has also been elucidated by our previous NMR experiment [24]. The relation between the pressure and the central frequency ($f$) of the NQR spectra of $Cu_2O$ powder at different temperatures ($T$) and pressures ($P$) has been determined in ref. [111] with the formula: $f(P,T) = A + BP + CP^2$, where $A = 27.06 - 0.4762\rho(T)$, $B = 0.4154 - 0.02682\rho(T)$, $C = -2.992 \times 10^{-4}\rho(T)$. In this formula, $\rho(T)$ is the $T$-dependent phonon distribution function, which is given by $\rho(T) = \frac{1}{2}\coth\left(\frac{h\nu_l}{2k_BT}\right)$, and $\nu_l$ is the phonon frequency, which is approximately $2.82 \times 10^{-4}$ MHz.

## 4. $^{51}$V NMR full spectra below $T_{CDW}$ at ambient pressure in $AV_3Sb_5$ (A = K, Rb and Cs).

In $AV_3Sb_5$ (A = K, Rb and Cs), the $^{51}$V NMR spectrum exhibited seven transition lines with the magnetic field applied along the c axis above $T_{CDW}$. Below $T_{CDW}$, as the kagome lattice reconstructs into a staggered tri-hexagonal superlattice, two inequivalent vanadium sites are generated within the hexagon cluster (V(I)) and the triangle cluster (V(II)). As shown in Fig. 8, the seven $^{51}$V NMR transition lines split into two sets of peaks with equal intensity.

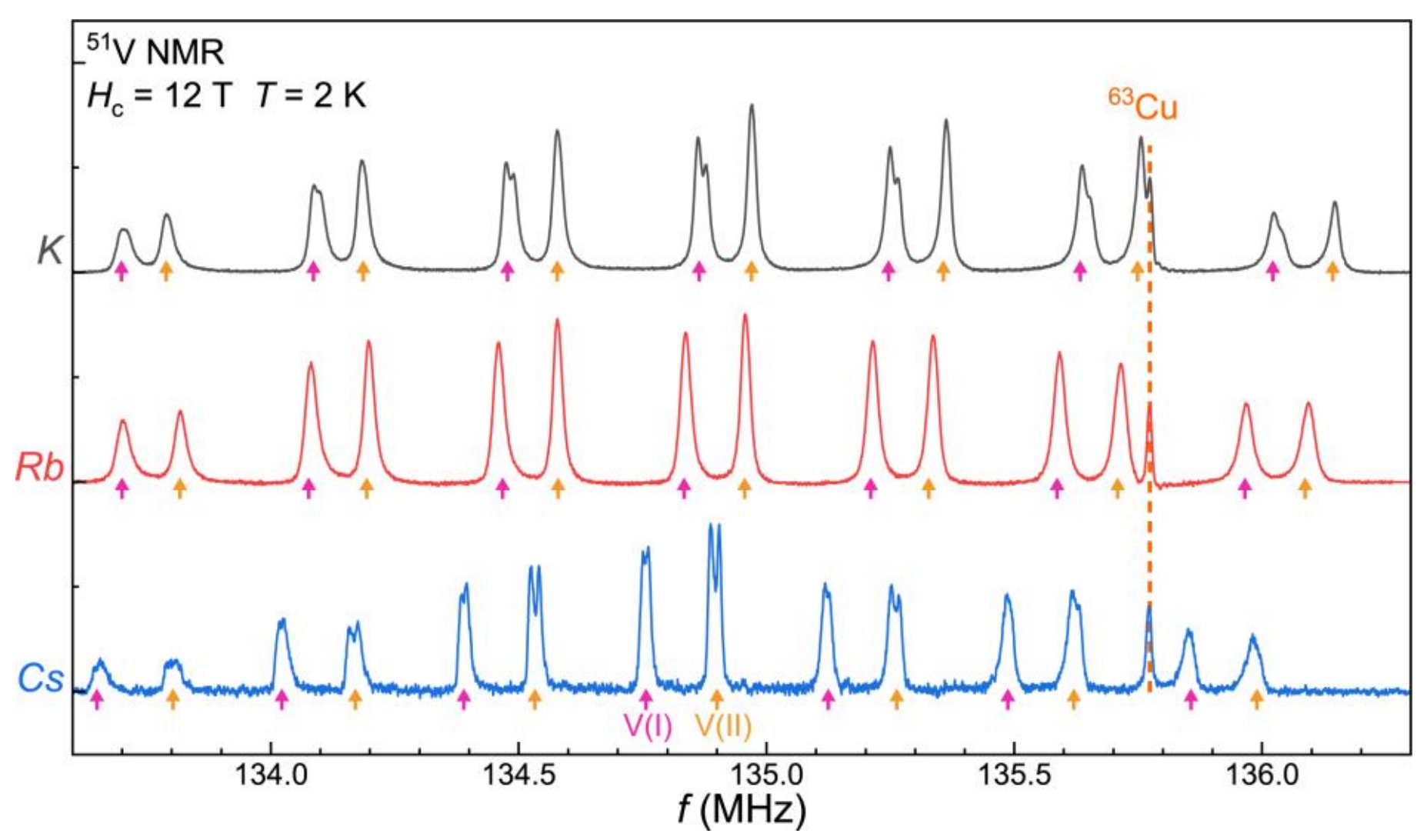

**FIG. 8 The typical $^{51}$V NMR full spectra of $AV_3Sb_5$(A = K, Rb and Cs) at ambient pressure below $T_{CDW}$.** The pink and orange arrows respectively mark for the vanadium sites in the hexagon cluster and the triangle cluster in **Fig. 1(b)**.

## 5. Pressure-dependent evolution of superconductivity in $AV_3Sb_5$ (A = K, Rb and Cs)

The NMR tank circuit can be used to detect superconducting transition by measuring the change in the resonant frequency of the NMR tank circuit. In our experiment, the NMR coil functions as a key inductive component within an $LC$ resonant circuit. When a sample inserted inside the coil cools down and enters the superconducting state, the diamagnetic shielding effect repels the magnetic flux, which effectively reduces the overall inductance of the NMR coil. As the resonant frequency of the $LC$ tank circuit is governed by the relationship $f = \frac{1}{2\pi\sqrt{LC}}$ , this physical reduction in inductance directly causes the resonant frequency to increase. Mathematically, this relative change in frequency $(-\Delta f/f)$ is directly proportional to the change in the sample's magnetic penetration depth $\Delta\lambda$. Consequently, by continuously monitoring this background-subtracted shift, the onset of the onset temperatures of superconducting transition is explicitly defined as the temperature where $-\Delta f/f$ deviates from zero.

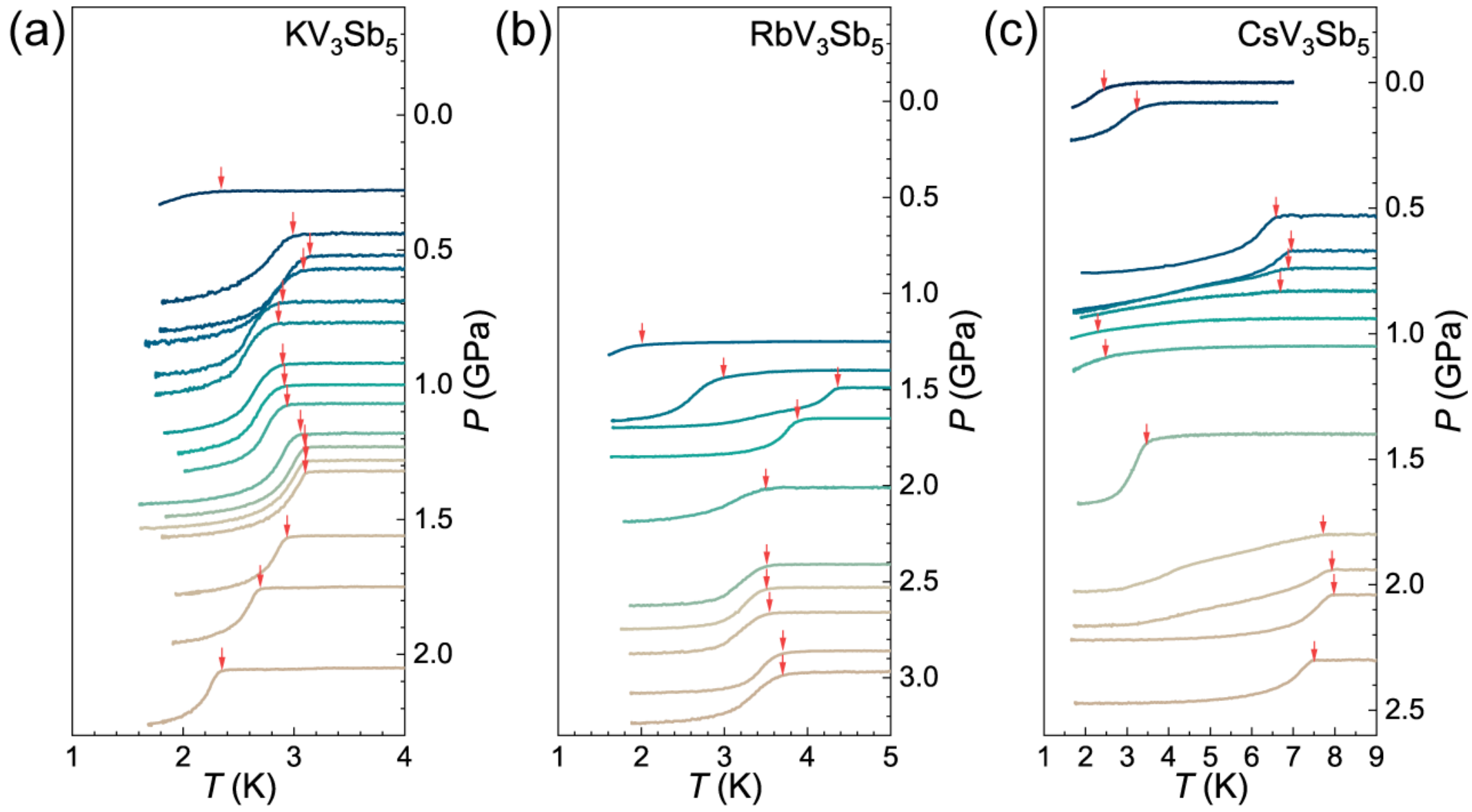

**FIG. 9 The *T*-dependent resonating frequency of the NMR tank circuit under different pressures in (a) $KV_3Sb_5$, (b) $RbV_3Sb_5$ and (c) $CsV_3Sb_5$.** The red arrows label the onset temperatures of superconducting transition, where the data and arrows in **(c)** are given by ref [24].

### 6. Simulation of the NMR spectra in the incommensurate CDW phases

Assuming that the relation between the NMR frequency shift and the nuclear displacements is purely local and linear [97], the triple-$Q$ CDW modulation at a V atom on the coordinate $\boldsymbol{r}$ can be denoted as:

$$\nu(\phi_1, \phi_2, \phi_3) = \nu_0 + \sum_{i=1,2,3} \nu_i \cos(\phi_i) = \nu_0 + \sum_{i=1,2,3} \nu_i \cos(\boldsymbol{q_i r} + \varphi_i).$$

Here $\boldsymbol{q_i}$ represents the various modulation wave vectors; $\nu_0$ is the center frequency; $\nu_i$ is the modulation amplitude, which is proportional to the CDW amplitude. As shown in Fig. 1(b), the standard staggered tri-hexagonal lattice is described using the ($MLL$) basis [73]:

$$\boldsymbol{q}_1^{(L)} = \left(\frac{1}{2}\ 0\ \frac{1}{2}\right), \qquad \boldsymbol{q}_2^{(L)} = \left(0\ \frac{1}{2}\ \frac{1}{2}\right), \qquad \boldsymbol{q}_3^{(M)} = \left(-\frac{1}{2}\ -\frac{1}{2}\ 0\right),$$

in the basis ($\boldsymbol{G}_1$, $\boldsymbol{G}_2$, $\boldsymbol{G}_3$):

$$\boldsymbol{G}_1 = \frac{2\pi}{a_0}\begin{bmatrix} 1 \\ 1/\sqrt{3} \\ 0 \end{bmatrix}, \boldsymbol{G}_2 = \frac{2\pi}{a_0}\begin{bmatrix} 0 \\ -2/\sqrt{3} \\ 0 \end{bmatrix}, \boldsymbol{G}_3 = \frac{2\pi}{c_0}\begin{bmatrix} 0 \\ 0 \\ 1 \end{bmatrix},$$

with $\nu_1 = \nu_2 \approx \nu_3$, and $\varphi_1 = 0$, $\varphi_2 = \frac{\pi}{2}, \varphi_3 = \frac{\pi}{2}$.

To simulate the incommensurate CDW modulation observed under pressure, the wave vectors are modified to $\boldsymbol{q}_i^* = \boldsymbol{q}_i(1-\delta)$, where $\delta$ constitutes the incommensurability parameter of the charge density waves. The simulation was performed on a kagome lattice containing 230400 ($12 \times 80 \times 80 \times 3$) vanadium atoms as shown in Fig. 10. We utilized $\delta = 0.005$, which provides sufficient scale to achieve stable incommensuration in the numerical results.

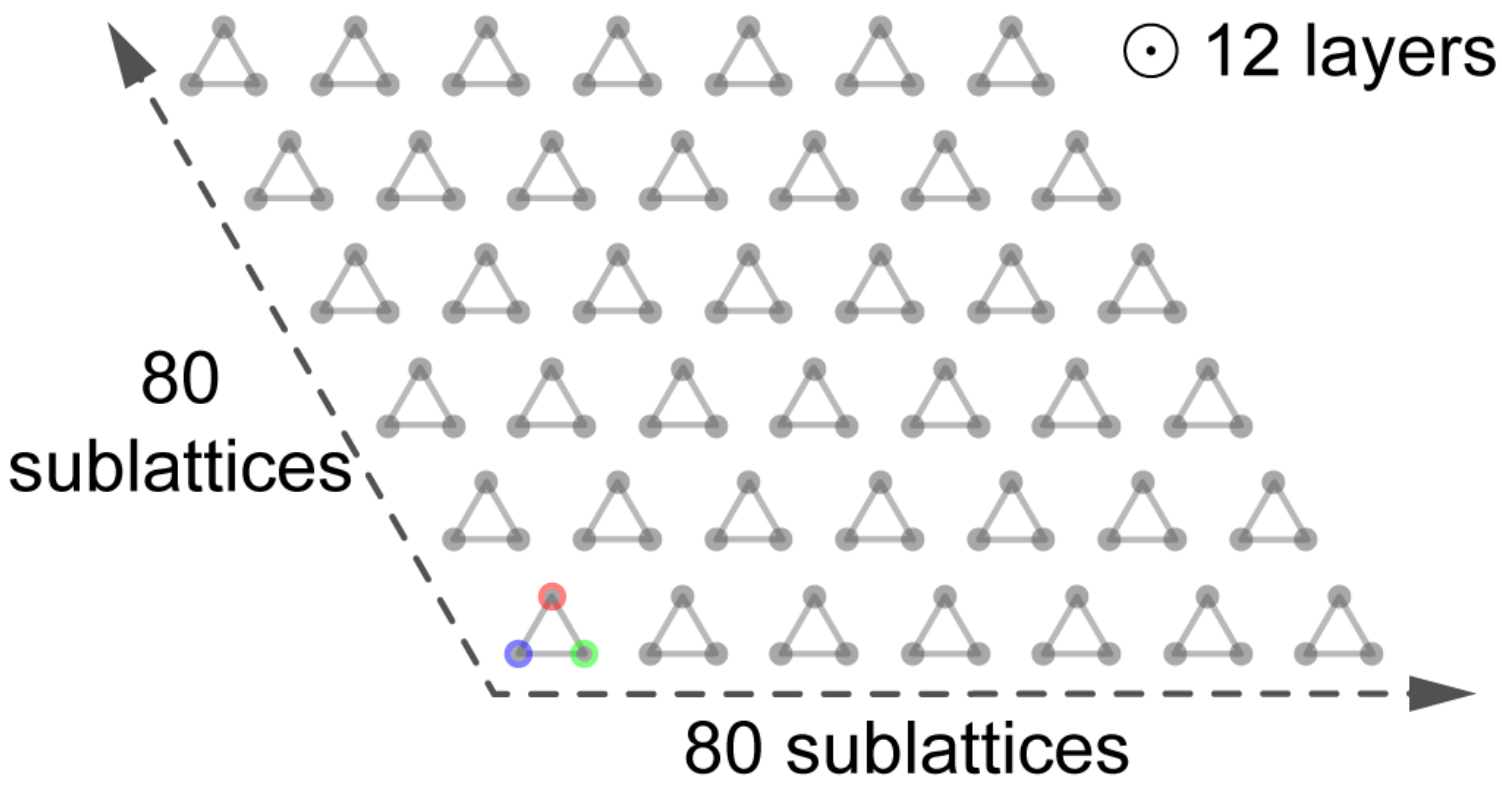

**FIG. 10 Sketch plot of the kagome lattice model in the simulation**. This model consists of 12 layers of vanadium atoms. Each layer contains an $80 \times 80$ grid of triangle sublattices with three V sites, which are highlighted in one sublattice with the three primary colors.

The spectral intensity $I(\nu)$ in the CDW phase is calculated by summing the frequency contributions of all $N$ atoms:

$$I(\nu) = \sum_j^N \left[\int \delta\left(\nu - \nu_j(\boldsymbol{r})\right) w(\nu)\, d\nu\right].$$

where $N = 230400$ and $w(\nu)$ is a Lorentzian broadening function:

$$w(\nu) = \frac{2A}{\pi}\frac{W}{4(\nu-\nu_0)^2 + W^2}.$$

where A is the intensity and W is the width of peak.

We first evaluated the single-$Q$ and double-$Q$ CDW models through spectral line shape simulations. As shown in Fig. 11(a), the line shape predicted by the single-$Q$ model fundamentally deviates from the experimental profile, manifesting as a deep spectral gap at the center. Although the double-$Q$ model with $\nu_1 = \nu_2$ does yield a central protrusion [Fig. 11(b)], its spectral weight is disproportionately pronounced compared to the subtle feature observed in $KV_3Sb_5$. Furthermore, under the condition $2\nu_1 = \nu_2$, the characteristic central plateau is entirely absent, replaced instead by a pronounced central depression. Because these simpler modulations fail to reproduce the key spectral features, we conclude that the experimental spectra are best described by the distinct triple-$Q$ incommensurate CDW modulations presented in Fig. 5.

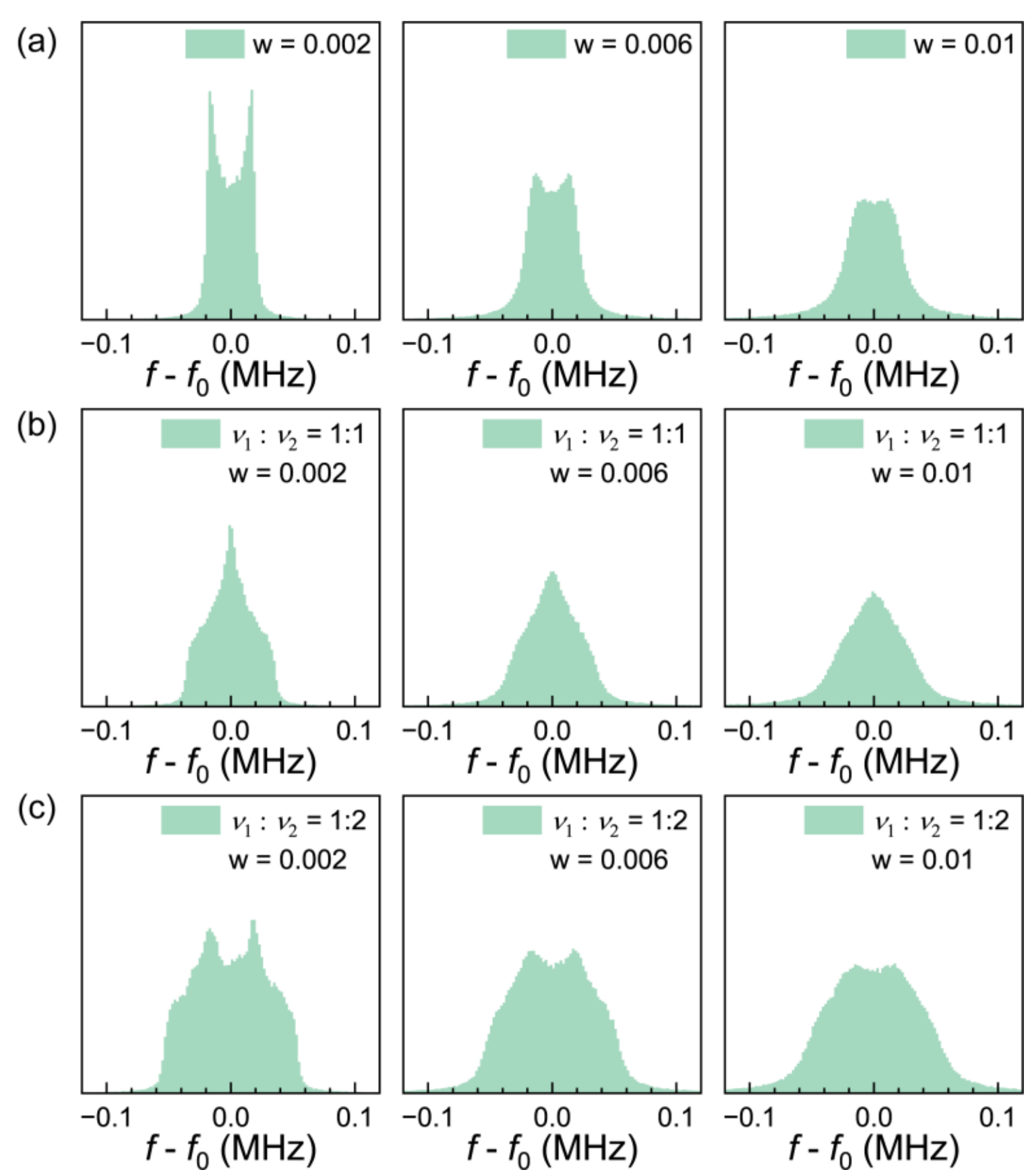

**FIG. 11 Single-$Q$ and double-$Q$ simulations of the $^{51}$V NMR spectra line shapes**. **(a)** Single-$Q$ incommensurate CDW with linewidth $w = 0, 0.006$ and 0.01 MHz. **(b)** Double-$Q$ incommensurate CDW with $\nu_1 = \nu_2$ and linewidth $w = 0, 0.006$ and 0.01 MHz. **(c)** Double-$Q$ incommensurate CDW with $\nu_1 = 2\nu_2$ and linewidth $w = 0, 0.006$ and 0.01 MHz.

## 7. First-principles calculations.

The temperature-dependent $N(E_F)$ is calculated with the formula $N(E_F) = \frac{1}{k_B T}\int(1 - f(E))f(E)D(E)dE$. Here, $k_B$ denotes the Boltzmann constant, $f(E)$ denotes the Fermi–Dirac distribution, and $D(E)$ denotes the DOS. The $T$-dependent $N(E_F)$ values are calculated with the vHs point located at different places relative to the Fermi surface in the kagome lattice. Here we take the density functional theory (DFT) calculation of the electronic structure of $CsV_3Sb_5$ at ambient pressure as example and adjust the alignment of vHSs at Fermi surface[24]. The DFT calculations were performed as implemented in the Vienna ab initio simulation package (VASP)[112].The phonon dispersions are obtained by using the PHONOPY code[113], and the results are shown in Fig. 12.

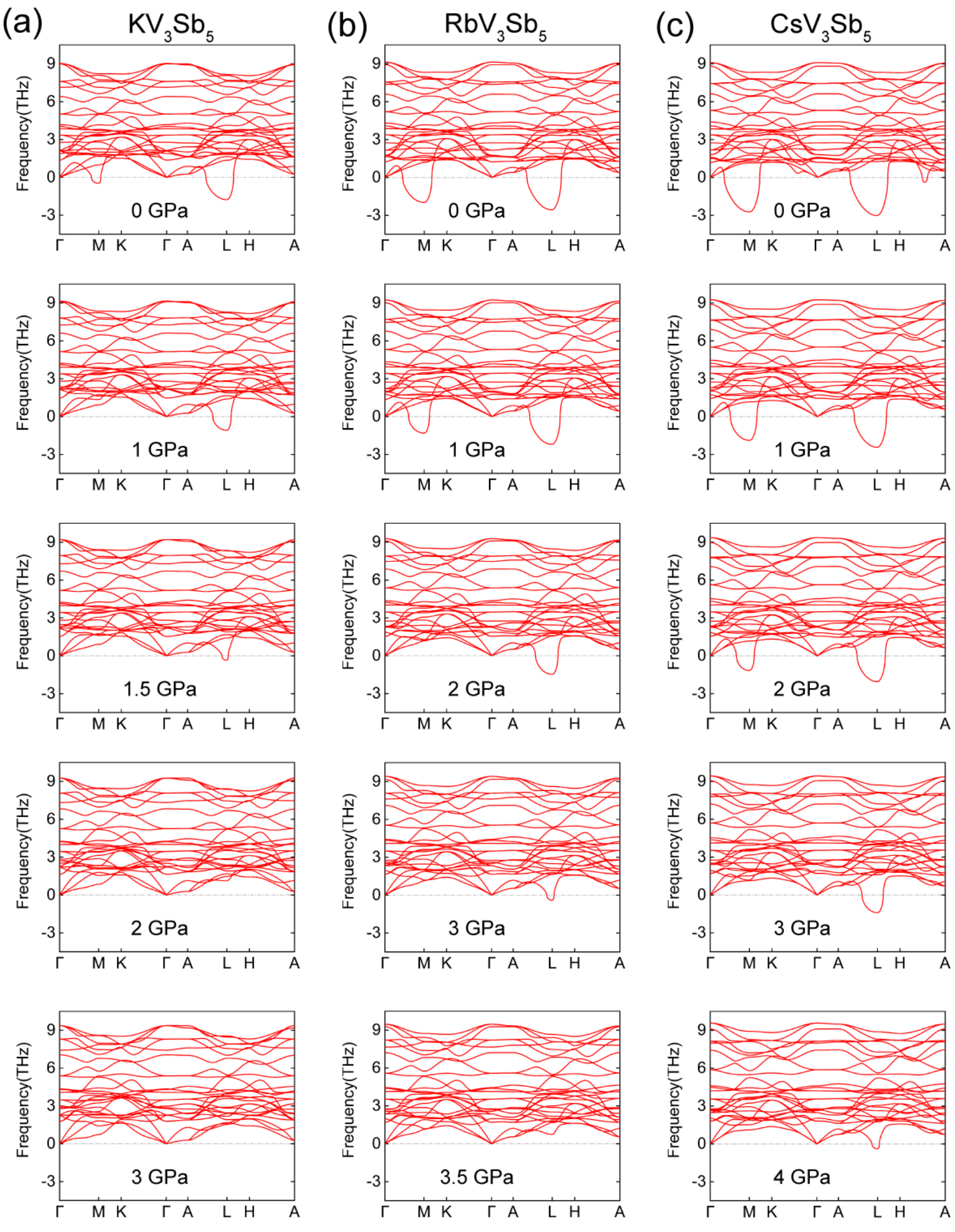

**FIG. 12 *P*-dependent phonon spectra in of (a) $KV_3Sb_5$, (b) $RbV_3Sb_5$ and (c) $CsV_3Sb_5$.** In $KV_3Sb_5$ and $RbV_3Sb_5$, the phonon imaginary frequency respectively persists up to 1.5 GPa and 3.0 GPa, which basically collapse into the critical pressure of the suppression of CDW. However, the phonon imaginary frequency persists up to 4.0 GPa in $CsV_3Sb_5$, which strikingly differs from the experimental result.

## Reference

[1] B. R. Ortiz, L. C. Gomes, J. R. Morey, M. Winiarski, M. Bordelon, J. S. Mangum, I. W. H. Oswald, J. A. RodriguezRivera, J. R. Neilson, S. D. Wilson *et al*., New kagome prototype materials: discovery of $KV_3Sb_5$, $RbV_3Sb_5$, and $CsV_3Sb_5$. Phys. Rev. Materials **3**, 094407 (2019).

[2] B. R. Ortiz, S. M. L. Teicher, Y. Hu, J. L. Zuo, P. M. Sarte, E. C. Schueller, A. M. Milinda Abeykoon, M. J. Krogstad, S. Rosenkranz, R. Osborn, R. Seshadri, L. Balents, J. He, and S. D. Wilson, $CsV_3Sb_5$ : A $Z_2$ Topological Kagome Metal with a Superconducting Ground State. Phys. Rev. Lett. **125**, 247002 (2020).

[3] B. R. Ortiz, P. M. Sarte, E. M. Kenney, M. J. Graf, S. M. L. Teicher, R. Seshadri, and S. D. Wilson, Superconductivity in the Z 2 kagome metal $KV_3Sb_5$. Phys. Rev. Materials **5**, 034801 (2021).

[4] Q. Yin, Z. Tu, C. Gong, Y. Fu, S. Yan, and H. Lei, Superconductivity and Normal-State Properties of Kagome Metal $RbV_3Sb_5$ Single Crystals. Chinese Phys. Lett. **38**, 037403 (2021).

[5] Y.-X. Jiang, *et al.* Unconventional chiral charge order in kagome superconductor $KV_3Sb_5$. Nat. Mater. **20**, 1353 (2021).

[6] N. Shumiya, *et al.* Intrinsic nature of chiral charge order in the kagome superconductor $RbV_3Sb_5$. Phys. Rev. B **104**, 035131 (2021).

[7] Z. Wang, *et al.* Electronic nature of chiral charge order in the kagome superconductor $CsV_3Sb_5$. Phys. Rev. B **104**, 075148 (2021).

[8] F. H. Yu, D. H. Ma, W. Z. Zhuo, S. Q. Liu, X. K. Wen, B. Lei, J. J. Ying, and X. H. Chen, Unusual competition of superconductivity and charge-density-wave state in a compressed topological kagome metal. Nat Commun **12**, 3645 (2021).

[9] C. Mielke, *et al.* Time-reversal symmetry-breaking charge order in a kagome superconductor. Nature **602**, 245 (2022).

[10] Y. Hu, *et al.* Time-reversal symmetry breaking in charge density wave of $CsV_3Sb_5$ detected by polar Kerr effect. arXiv:2208.08036 (2022).

[11] Y. Xu, Z. Ni, Y. Liu, B. R. Ortiz, Q. Deng, S. D. Wilson, B. Yan, L. Balents, and L. Wu, Three-state nematicity and magneto-optical Kerr effect in the charge density waves in kagome superconductors, Nat. Phys. **18**, 1470 (2022).

[12] Z. Guguchia, *et al.* Tunable unconventional kagome superconductivity in charge ordered $RbV_3Sb_5$ and $KV_3Sb_5$. Nat Commun **14**, 153 (2023).

[13] W. Liège, Y. Xie, D. Bounoua, Y. Sidis, F. Bourdarot, Y. Li, Z. Wang, J.-X. Yin, P. Dai, and P. Bourges, Search for orbital magnetism in the kagome superconductor $CsV_3Sb_5$ using neutron diffraction, Phys. Rev. B **110**, 195109 (2024).

[14] H. Deng, H. Qin, G. Liu, *et al.* Chiral kagome superconductivity modulations with residual Fermi arcs. Nature **632**, 775–781 (2024)

[15] H. Gui, L. Yang, X. Wang, *et al.* Probing orbital magnetism of a kagome metal $CsV_3Sb_5$ by a tuning fork resonator. Nat Commun **16**, 4275 (2025).

[16] Y. Xiang, Q. Li, Y. Li, W. Xie, H. Yang, Z. Wang, Y. Yao, and H.-H. Wen, Twofold symmetry of c-axis resistivity in topological kagome superconductor $CsV_3Sb_5$ with in-plane rotating magnetic field. Nature Communications **12**, 6727 (2021).

[17] H. Zhao, H. Li, B. R. Ortiz, S. M. L. Teicher, T. Park, M. Ye, Z. Wang, L. Balents, S. D. Wilson, and I. Zeljkovic, Cascade of correlated electron states in the kagome superconductor $CsV_3Sb_5$. Nature **599**, 216 (2021).

[18] H. Chen *et al.* Roton pair density wave in a strong-coupling kagome superconductor. Nature **599**, 222 (2021).

[19] L. Nie *et al.* Charge-density-wave-driven electronic nematicity in a kagome superconductor. Nature **604**, 59 (2022).

[20] Y. Xu, Z. Ni, Y. Liu, B. R. Ortiz, Q. Deng, S. D. Wilson, B. Yan, L. Balents, and L. Wu, Three-state nematicity and magneto-optical Kerr effect in the charge density waves in kagome superconductors. Nat. Phys. **18**, 1470 (2022).

[21] H. Li, H. Zhao, B. R. Ortiz, Y. Oey, Z. Wang, S. D. Wilson, and I. Zeljkovic, Unidirectional coherent quasiparticles in the high-temperature rotational symmetry broken phase of $AV_3Sb_5$ kagome superconductors. Nat. Phys. **19**, 637 (2023).

[22] F. Du, S. Luo, B. R. Ortiz, Y. Chen, W. Duan, D. Zhang, X. Lu, S. D. Wilson, Y. Song, and H. Yuan, Pressure-induced double superconducting domes and charge instability in the kagome metal $KV_3Sb_5$. Phys. Rev. B **103**, L220504 (2021).

[23] N. N. Wang *et al.* Competition between charge-density-wave and superconductivity in the kagome metal $RbV_3Sb_5$. Phys. Rev. Research **3**, 043018 (2021).

[24] L. Zheng *et al*. Emergent charge order and unconventional superconductivity in pressurized kagome superconductor $CsV_3Sb_5$. Nature **611**, 682–687 (2022).

[25] R. Gupta, D. Das, C. Mielke, *et al.* Two types of charge order with distinct interplay with superconductivity in the kagome material $CsV_3Sb_5$. Commun Phys **5**, 232 (2022).

[26] X. Wen et al., Emergent superconducting fluctuations in compressed kagome superconductor $CsV_3Sb_5$, Science Bulletin **68**, 259 (2023).

[27] M. Kang, S. Fang, J. Yoo, *et al.* Charge order landscape and competition with superconductivity in kagome metals. Nat. Mater. **22**, 186–193 (2023)

[28] M. Roppongi, K. Ishihara, Y. Tanaka, *et al.* Bulk evidence of anisotropic *s*-wave pairing with no sign change in the kagome superconductor $CsV_3Sb_5$. Nat Commun **14**, 667 (2023)

[29] X.-Y. Yan *et al.* Chiral 2×2 pair density waves with residual Fermi arcs in $RbV_3Sb_5$, Chinese Phys. Lett. **41**, 097401 (2024).

[30] F. Stier et al., Pressure-Dependent Electronic Superlattice in the Kagome Superconductor $CsV_3Sb_5$. Phys. Rev. Lett. **133**, 236503 (2024).

[31] Y. Sun et al., Imaging momentum-space Cooper pair formation and its competition with the charge density wave gap in a kagome superconductor. Sci. China Phys. Mech. Astron. **67**, 277411 (2024).

[32] H. Deng, H. Qin, Liu, G. *et al.* Chiral kagome superconductivity modulations with residual Fermi arcs. Nature **632**, 775–781 (2024)

[33] L. Huai *et al.* Electron-Correlation-Assisted Charge Stripe Order in a Kagome Superconductor, Phys. Rev. X **15**, 041039 (2025).

[34] T. Nagashima, K. Ishihara, Y. Yamakawa, *et al.* Impact of charge-density-wave pattern on the superconducting gap in Vanadium-based kagome superconductors. Commun Phys **8**, 303 (2025)

[35] X. Han, H. Chen, H. Tan, *et al.* Atomic manipulation of the emergent quasi-2D superconductivity and pair density wave in a kagome metal. Nat. Nanotechnol. **20**, 1017–1025 (2025)

[36] M. L. Kiesel, C. Platt, and R. Thomale, Unconventional Fermi Surface Instabilities in the Kagome Hubbard Model. Phys. Rev. Lett. **110**, 126405 (2013).

[37] H. Li, *et al.* Observation of unconventional charge density wave without acoustic phonon anomaly in kagome superconductors $AV_3Sb_5$ (A=Rb, Cs). Phys. Rev. X **11**, 031050 (2021).

[38] Z. Wang *et al.* Distinctive momentum dependent charge-density-wave gap observed in $CsV_3Sb_5$ superconductor with topological kagome lattice, arXiv:2104.05556 (2021).

[39] S. Cho *et al.* Emergence of New van Hove Singularities in the Charge Density Wave State of a Topological Kagome Metal $RbV_3Sb_5$. Phys. Rev. Lett. **127**, 236401 (2021).

[40] H. Tan, Y. Liu, Z. Wang, and B. Yan, Charge Density Waves and Electronic Properties of Superconducting Kagome Metals. Phys. Rev. Lett. **127**, 046401 (2021).

[41] M. M. Denner, R. Thomale, and T. Neupert, Analysis of Charge Order in the Kagome Metal $AV_3Sb_5$ (A = K, Rb, Cs). Phys. Rev. Lett. **127**, 217601 (2021).

[42] X. Zhou, Y. Li, X. Fan, J. Hao, Y. Dai, Z. Wang, Y. Yao, and H.-H. Wen, Origin of the Charge Density Wave in the Kagome Metal $CsV_3Sb_5$ as Revealed by Optical Spectroscopy. Phys. Rev. B **104**, L041101 (2021).

[43] X. Feng, K. Jiang, Z. Wang, and J. Hu, Chiral flux phase in the Kagome superconductor $AV_3Sb_5$, Science Bulletin 66, 1384 (2021).

[44] H. Li, S. Wan, H. Li, Q. Li, Q. Gu, H. Yang, Y. Li, Z. Wang, Y. Yao, and H.-H. Wen, No observation of chiral flux current in the topological kagome metal $CsV_3Sb_5$. Phys. Rev. B **105**, 045102 (2022).

[45] C. Guo *et al.* Switchable chiral transport in charge-ordered kagome metal $CsV_3Sb_5$. Nature **611**, 461 (2022).

[46] T. Neupert, M. M. Denner, J. X. Yin, *et al.* Charge order and superconductivity in kagome materials. Nat. Phys. **18**, 137–143 (2022)

[47] Y. Hu *et al.*, Topological surface states and flat bands in the kagome superconductor $CsV_3Sb_5$. Science Bulletin **67**, 495 (2022).

[48] M. Kang *et al.* Twofold van Hove singularity and origin of charge order in topological kagome superconductor $CsV_3Sb_5$. Nat. Phys. **18**, 301 (2022).

[49] T. Kato *et al.* Three-dimensional energy gap and origin of charge-density wave in kagome superconductor $KV_3Sb_5$. Commun Mater **3**, 30 (2022).

[50] H. Luo *et al.* Electronic nature of charge density wave and electron-phonon coupling in kagome superconductor $KV_3Sb_5$. Nat Commun **13**, 273 (2022).

[51] J.-T. Jin, K. Jiang, H. Yao, and Y. Zhou, Interplay between Pair Density Wave and a Nested Fermi Surface, Phys. Rev. Lett. 129, 167001 (2022).

[52] F. Kaboudvand, S. M. L. Teicher, S. D. Wilson, R. Seshadri, and M. D. Johannes, Fermi surface nesting and the Lindhard response function in the kagome superconductor $CsV_3Sb_5$, Applied Physics Letters **120**, 111901 (2022).

[53] E. Uykur, B. R. Ortiz, S. D. Wilson, M. Dressel, and A. A. Tsirlin, Optical detection of the density-wave instability in the kagome metal $KV_3Sb_5$, Npj Quantum Mater. **7**, 16 (2022).

[54] S. Wu, B. R. Ortiz, H. Tan, S. D. Wilson, B. Yan, T. Birol, and G. Blumberg, Charge density wave order in the kagome metal $AV_3Sb_5$ (A = Cs, Rb, K), Phys. Rev. B **105**, 155106 (2022).

[55] D. R. Saykin et al., High Resolution Polar Kerr Effect Studies of $CsV_3Sb_5$: Tests for Time-Reversal Symmetry Breaking below the Charge-Order Transition. Phys. Rev. Lett. **131**, 016901 (2023).

[56] D. Subires, A. Korshunov, A.H. Said, *et al.* Order-disorder charge density wave instability in the kagome metal $(Cs,Rb)V_3Sb_5$. Nat Commun **14**, 1015 (2023)

[57] Y. Zhong *et al.* Testing electron–phonon coupling for the superconductivity in kagome metal $CsV_3Sb_5$. Nat Commun **14**, 1945 (2023).

[58] Y. Zhong, *et al.* Nodeless electron pairing in $CsV_3Sb_5$-derived kagome superconductors. Nature **617**, 488–492 (2023).

[59] D. Azoury *et al.* Direct observation of the collective modes of the charge density wave in the kagome metal $CsV_3$ $Sb_5$, Proc. Natl. Acad. Sci. U.S.A. **120**, e2308588120 (2023).

[60] J. Deng, R. Zhang, Y. Xie, X. Wu, and Z. Wang, Two elementary band representation model, Fermi surface nesting, and surface topological superconductivity in $AV_3Sb_5$ (A = K, Rb, Cs), Phys. Rev. B **108**, 115123 (2023).

[61] Z. Liu *et al.* Absence of $E_{2g}$ Nematic Instability and Dominant $A_{1g}$ Response in the Kagome Metal $CsV_3Sb_5$. Phys. Rev. X **14**, 031015 (2024).

[62] S.D. Wilson, B.R. Ortiz, $AV_3Sb_5$ kagome superconductors. Nat Rev Mater **9**, 420–432 (2024)

[63] G. He *et al.* Anharmonic strong-coupling effects at the origin of the charge density wave in $CsV_3Sb_5$. Nat Commun **15**, 1895 (2024).

[64] M. Gutierrez-Amigo, Đ. Dangić, C. Guo, C. Felser, P. J. W. Moll, M. G. Vergniory, and I. Errea, Phonon collapse and anharmonic melting of the 3D charge-density wave in kagome metals, Commun Mater **5**, 234 (2024).

[65] X. Wu, D. Chakraborty, A. P. Schnyder, and A. Greco, Crossover between electron-electron and electron-phonon mediated pairing on the kagome lattice, Phys. Rev. B **109**, 014517 (2024).

[66] J.-Y. You, C.-E. Hsu, M. Del Ben, and Z. Li, Diverse Manifestations of Electron-Phonon Coupling in a Kagome Superconductor, Phys. Rev. Lett. **134**, 106401 (2025).

[67] M. Alkorta, M. Gutierrez-Amigo, Đ. Dangić, C. M. Guo, P. J. W. Moll, M. G. Vergniory, and I. Errea, Symmetry-Broken Ground State and Phonon-Mediated Superconductivity in Kagome $CsV_3Sb_5$, Phys. Rev. Lett. **136**, 206401 (2026).

[68] P. H. McGuinness *et al.* Soft mode origin of charge ordering in superconducting kagome $CsV_3Sb_5$, Nat Commun **17**, 4817 (2026).

[69] X. Yan, G. Liu, H. Deng, X. Xu, H. Ma, H. Qin, J. Zhang, Y. Zhao, X. Fan, W. Song, M. Gao, H. Zhao, Z. Qu, Y. Zhong, K. Okazaki, X. Zheng, Y. Peng, Z. Guguchia, X. Wu, D. Wang, Q. Wang, H. Hohmann, M. Dürrnagel, R. Thomale, & J. Yin, Phase-sensitive evidence for pair density waves in a kagome superconductor, Proc. Natl. Acad. Sci. **123** (22) e2604142123,

[70] Y. Lan, Y. Lei, C. Le, B. R. Ortiz, N. C. Plumb, M. Radovic, X. Wu, M. Shi, S. D. Wilson, and Y. Hu, Common Sublattice-Pure Van Hove Singularities in the Kagome Superconductors $AV_3Sb_5$ (A = K , Rb, Cs). Phys. Rev. Lett. **136**, 016401 (2026).

[71] W.-S. Wang, Z.-Z. Li, Y.-Y. Xiang, and Q.-H. Wang, Competing electronic orders on kagome lattices at van Hove filling. Phys. Rev. B **87**, 115135 (2013).

[72] Y.-P. Lin and R. M. Nandkishore, Complex charge density waves at Van Hove singularity on hexagonal lattices: Haldane-model phase diagram and potential realization in the kagome metals $AV_3Sb_5$ ( A =K, Rb, Cs). Phys. Rev. B **104**, 045122 (2021).

[73] M. H. Christensen, T. Birol, B. M. Andersen, and R. M. Fernandes, Theory of the charge density wave in $AV_3Sb_5$ kagome metals. Phys. Rev. B **104**, 214513 (2021).

[74] X. Wu *et al.* Nature of Unconventional Pairing in the Kagome Superconductors $AV_3Sb_5$ (A = K, Rb, Cs). Phys. Rev. Lett. **127**, 177001 (2021).

[75] M. Y. Jeong, H.-J. Yang, H. S. Kim, Y. B. Kim, S. Lee, and M. J. Han, Crucial role of out-of-plane Sb p orbitals in Van Hove singularity formation and electronic correlations in the superconducting kagome metal $CsV_3Sb_5$. Phys. Rev. B **105**, 235145 (2022).

[76] H. Li, Y. B. Kim, and H.-Y. Kee, Intertwined Van Hove Singularities as a Mechanism for Loop Current Order in Kagome Metals. Phys. Rev. Lett. **132**, 146501 (2024).

[77] Y. Tian and S. Y. Savrasov, Unconventional superconductivity via charge fluctuations in the kagome metal $CsV_3Sb_5$. Phys. Rev. B **111**, 134507 (2025).

[78] N. Ratcliff, L. Hallett, B. R. Ortiz, S. D. Wilson, and J. W. Harter, Coherent phonon spectroscopy and interlayer modulation of charge density wave order in the kagome metal $CsV_3Sb_5$, Phys. Rev. Materials 5, L111801 (2021).

[79] Z. Ye, A. Luo, J.-X. Yin, M. Z. Hasan, and G. Xu, Structural instability and charge modulations in the kagome superconductor $AV_3Sb_5$. Phys. Rev. B **105**, 245121 (2022).

[80] E. T. Ritz, R. M. Fernandes, and T. Birol, Impact of Sb degrees of freedom on the charge density wave phase diagram of the kagome metal $CsV_3Sb_5$, Phys. Rev. B **107**, 205131 (2023).

[81] G. Liu *et al.* Observation of anomalous amplitude modes in the kagome metal $CsV_3Sb_5$. Nat Commun **13**, 3461 (2022).

[82] P. Wu *et al.* Unidirectional electron–phonon coupling in the nematic state of a kagome superconductor. Nat. Phys. **19**, 1143–1149 (2023).

[83] Y. Xie *et al.* Electron-phonon coupling in the charge density wave state of $CsV_3Sb_5$. Phys. Rev. B **105**, L140501 (2022).

[84] M. Wenzel, B. R. Ortiz, S. D. Wilson, M. Dressel, A. A. Tsirlin, and E. Uykur, Optical study of $RbV_3Sb_5$: Multiple density-wave gaps and phonon anomalies. Phys. Rev. B **105**, 245123 (2022).

[85] Y. Wang *et al.* Soft Phonon Charge-Density-Wave Formation in the Kagome Metal $KV_3Sb_5$, Phys. Rev. Lett. **136**, 136401 (2026).

[86] H.-S. Xu, Y.-J. Yan, R. Yin, W. Xia, S. Fang, Z. Chen, Y. Li, W. Yang, Y. Guo, and D.-L. Feng, Multiband Superconductivity with Sign-Preserving Order Parameter in Kagome Superconductor $CsV_3Sb_5$. Phys. Rev. Lett. **127**, 187004 (2021).

[87] W. Duan *et al.* Nodeless superconductivity in the kagome metal $CsV_3Sb_5$. Science China Physics, Mechanics & Astronomy **64**, 107462 (2021).

[88] C. Mu, Q. Yin, Z. Tu, C. Gong, H. Lei, Z. Li, and J. Luo,S-Wave Superconductivity in Kagome Metal $CsV_3Sb_5$ Revealed by $^{121/123}$Sb NQR and $^{51}$V NMR Measurements. Chinese Phys. Lett. **38**, 077402 (2021).

[89] X. Y. Feng, Z. Zhao, J. Luo, Y. Z. Zhou, J. Yang, A. F. Fang, H. T. Yang, H.-J. Gao, R. Zhou, and G. Zheng, Fully-gapped superconductivity with rotational symmetry breaking in pressurized kagome metal $CsV_3Sb_5$. Nat Commun **16**, 3643 (2025).

[90] H. LaBollita and A. S. Botana, Tuning the Van Hove singularities in $AV_3Sb_5$ (A = K, Rb, Cs) via pressure and doping. Phys. Rev. B **104**, 205129 (2021).

[91] S. Sim, M. Y. Jeong, H. Lee, D. H. D. Lee, and M. J. Han, Chemical effect on the Van Hove singularity in superconducting kagome metal $AV_3Sb_5$ (A = K, Rb, Cs). Phys. Chem. Chem. Phys. **26**, 11715–11721 (2024).

[92] X. Zhou, Y. Li, X. Fan, J. Hao, Y. Xiang, Z. Liu, Y. Dai, Z. Wang, Y. Yao, and H.-H. Wen, Electronic correlations and evolution of the charge density wave in the kagome metals $AV_3Sb_5$ (A = K, Rb, Cs). Phys. Rev. B **107**, 165123 (2023).

[93] D. Song *et al.* Orbital ordering and fluctuations in a kagome superconductor $CsV_3Sb_5$. Sci. China Phys. Mech. Astron. **65**, 247462 (2022).

[94] L. Wang *et al*. Giant critical current peak induced by pressure in kagome superconductor $RbV_3Sb_5$, arXiv:2511.21195 (2025).

[95] Z. Wang *et al*. Discovery of a New Phase in Thin Flakes of $KV_3Sb_5$ under Pressure, Advanced Science 12, 2415012 (2025).

[96] See Supplementary Material for additional data and analysis.

[97] R. Blinc, S. Žumer, NMR line shapes and relaxation in incommensurate systems with a multiple-q modulation. Phys. Rev. B **16**, 11314 (1990).

[98] Y. I. Joe *et al.* Emergence of charge density wave domain walls above the superconducting dome in 1*T*-$TiSe_2$. Nature Phys **10**, 421–425 (2014).

[99] A. Kogar *et al.* Observation of a Charge Density Wave Incommensuration Near the Superconducting Dome in $Cu_xTiSe_2$. Phys. Rev. Lett. **118**, 027002 (2017).

[100] D. Novko, Z. Torbatian, and I. Lončarić, Electron correlations rule the phonon-driven instability in single-layer $TiSe_2$, Phys. Rev. B **106**, 245108 (2022)

[101] J.-W. Dong, Z. Wang, and S. Zhou, Loop-current charge density wave driven by long-range Coulomb repulsion on the kagomé lattice, Phys. Rev. B **107**, 045127 (2023).

[102] K. Zeng, Z. Wang, K. Jiang, & Z. Wang, Electronic structure of $AV_3Sb_5$ kagome metals. Phys. Rev. B **111**, 235114 (2025).

[103] J. Zhan *et al.* Loop current order on the kagome lattice. Phys. Rev. Lett. **136**, 126001 (2026).

[104] Q. Li, M. Hücker, G. D. Gu, A. M. Tsvelik, & J. M. Tranquada, Two-dimensional superconducting fluctuations in stripe-ordered $La_{1.875}Ba_{0.125}CuO_4$. Phys. Rev. Lett. **99**, 067001 (2007).

[105] M. H. Christensen, T. Birol, B. M. Andersen, and R. M. Fernandes, Loop currents in $AV_3Sb_5$ kagome metals: Multipolar and toroidal magnetic orders, Phys. Rev. B 1**4**, 144504 (2022).

[106] Z. Wang, K. Zeng, and Z. Wang, Roton Superconductivity from Loop-Current Chern Metal on the Kagome Lattice, arXiv:2504.02751(2025).

[107] R. M. Fernandes, T. Birol, M. Ye, and D. Vanderbilt, Loop-current order through the kagome looking glass, arXiv:2502.16657 (2025).

[108] H. D. Scammell, J. Ingham, T. Li, and O. P. Sushkov, Chiral excitonic order from twofold van Hove singularities in kagome metals. Nat Commun **14**, 605 (2023).

[109] W. W. Simmons, W. J. O'Sullivan, and W. A. Robinson, Nuclear Spin-Lattice Relaxation in Dilute Paramagnetic Sapphire. Phys. Rev. **127**, 1168–1178 (1962).

[110] D. E. MacLaughlin, J. D. Williamson, and J. Butterworth, J. Nuclear Spin-Lattice Relaxation in Pure and Impure Indium. I. Normal State. Phys. Rev. B **4**, 60–70 (1971).

[111] K. Kitagawa, H. Gotou, T. Yagi, A. Yamada, T. Matsumoto, Y. Uwatoko, and M. Takigawa, Space efficient opposed-anvil high-pressure cell and its application to optical and NMR measurements up to 9GPa. J. Phys. Soc. Jpn **79**, 024001 (2010).

[112] G. Kresse, & J. Furthmüller, Efficient iterative schemes for ab initio total-energy calculations using a plane-wave basis set. Phys. Rev. B. **54**, 11169 (1996).

[113] A. Togo and I. Tanaka, First principles phonon calculations in materials science, Scripta Materialia **108**, 1 (2015).

# Supplementary Materials for

# "Pressure-induced unconventional charge-density-wave states in kagome metal $AV_3Sb_5$ (A = K, Rb, Cs)"

Zhimian Wu[1], Linpeng Nie[1], Ye Yang[3], Kuanglv Sun[1], Huachen Rao[1], Dan Zhao[1], Zhongjun Li[3], Tao Wu[1,2,4,5,*] and Xianhui Chen[1,2,4,5,†]

1. Hefei National Research Center for Physical Sciences at the Microscale, University of Science and Technology of China, Hefei, Anhui 230026, China
2. Department of Physics, University of Science and Technology of China, Hefei, Anhui 230026, China
3. School of Physics, Hefei University of Technology, Hefei 230009, China
4. Collaborative Innovation Center of Advanced Microstructures, Nanjing University, Nanjing 210093, China
5. Hefei National Laboratory, University of Science and Technology of China, Hefei 230088, China

*wutao@ustc.edu.cn
†chenxh@ustc.edu.cn

**Outline**

**Section I. Hydrostatic pressure calibration.**

**Section II. Temperature-dependent $^{51}$V NMR spectra at different pressures in $AV_3Sb_5$ (A = K, Rb).**

**Section III. Temperature-dependent $^{51}$V NMR $1/T_1T$ at different pressures in $AV_3Sb_5$ (A = K, Rb).**

**Section IV. Typical recovery curves for $T_1$ and temperature dependence of parameter $\beta$ in $T_1$ fitting at different pressures in $AV_3Sb_5$ (A = K, Rb).**

**Section V. Temperature-dependent $^{51}$V NMR central transition lines at $P_{c2}$ in $AV_3Sb_5$ (A = K, Rb and Cs).**

## Section I. Hydrostatic pressure calibration.

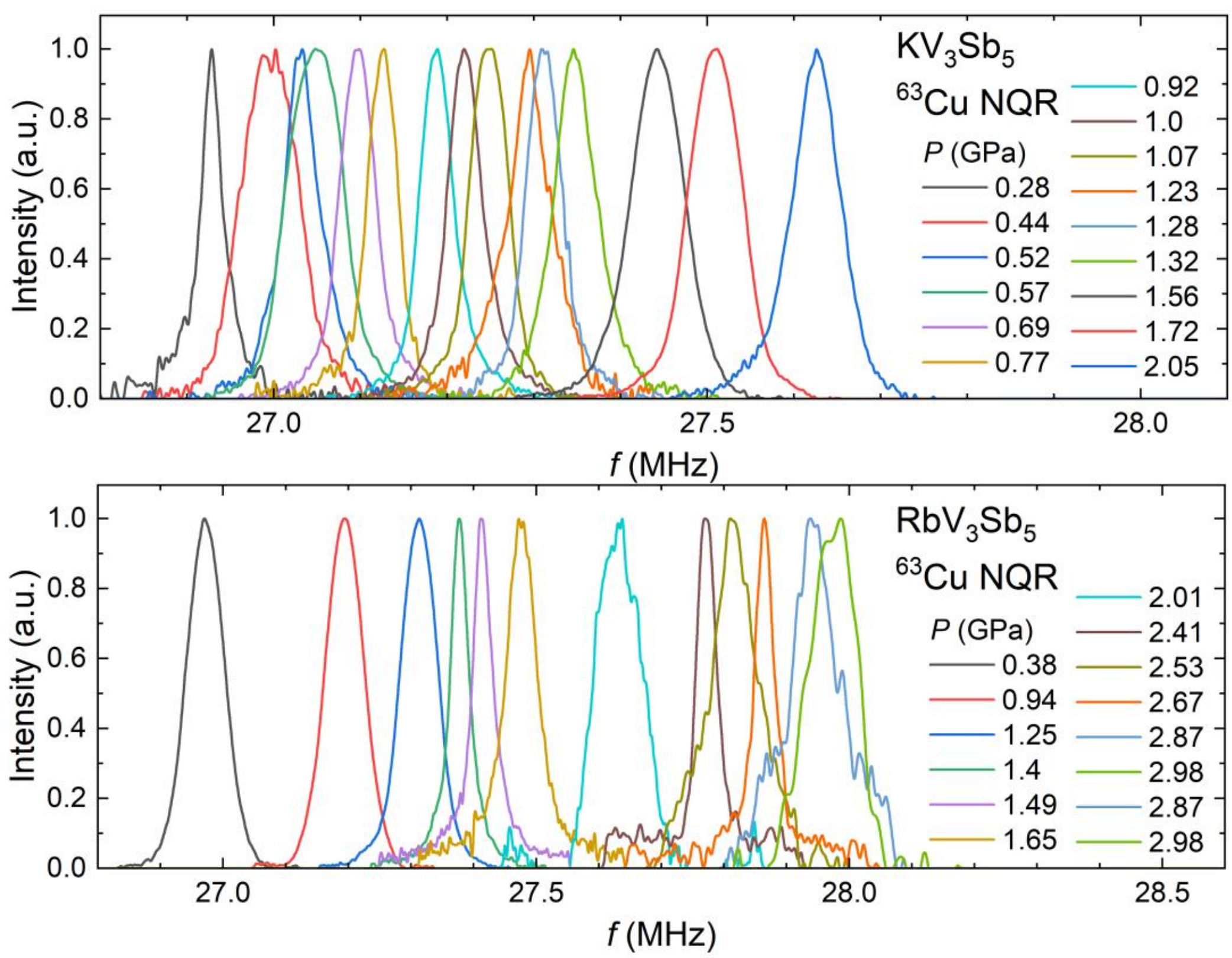

**FIG. S1 $^{63}$Cu NQR spectra of $Cu_2O$ at *T* = 2 K.** The applied hydrostatic pressure of $KV_3Sb_5$ and $RbV_3Sb_5$ were calibrated using the pressure dependent frequency at the peak of $^{63}$Cu NQR spectra in $Cu_2O$ powder measured at 2 K.

## Section II. Temperature-dependent $^{51}$V NMR central transition lines under different pressures in $AV_3Sb_5$ (A = K, Rb).

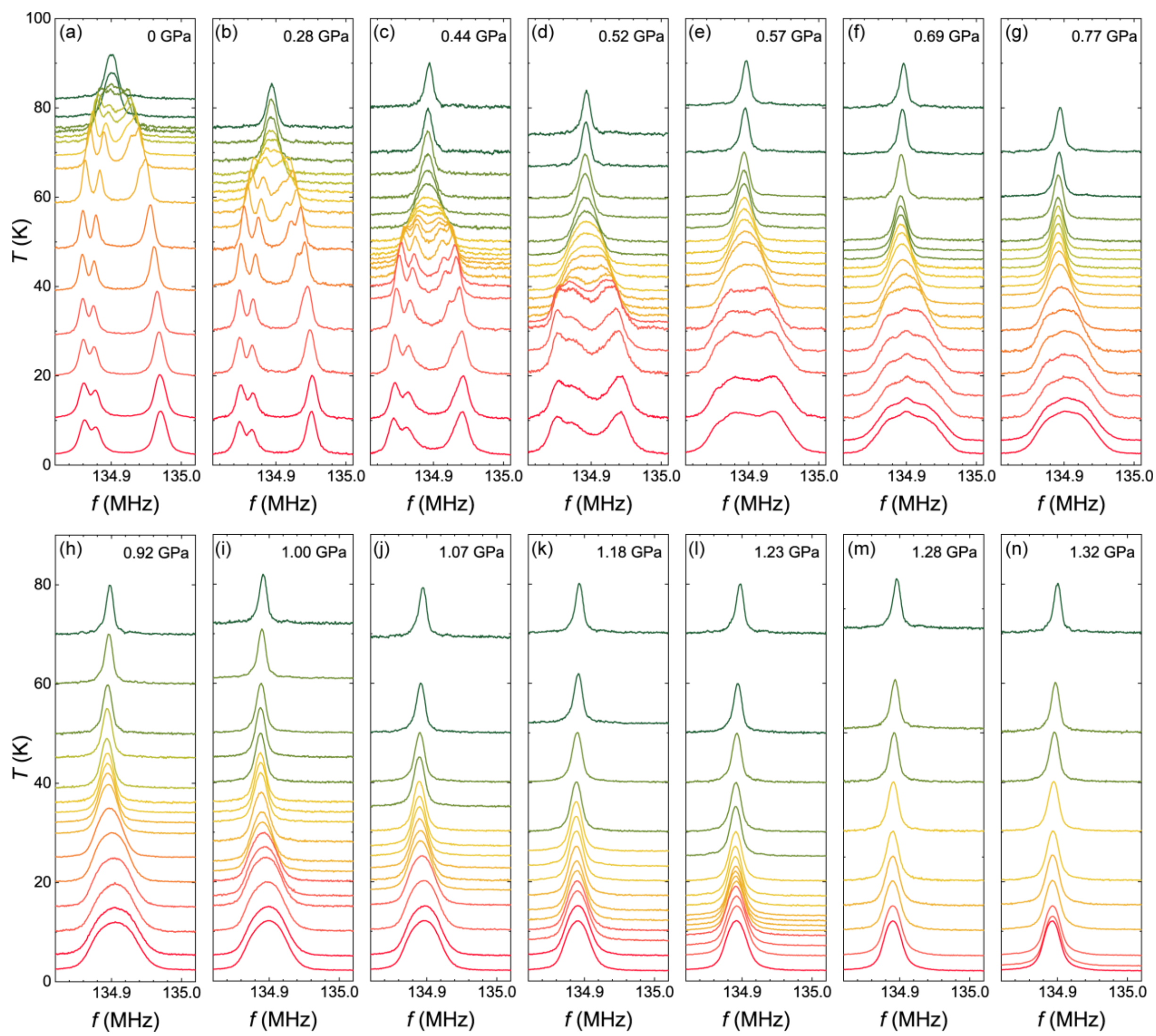

**FIG. S2 *T*-dependent $^{51}$V NMR spectra of $KV_3Sb_5$ at different pressures.** (a)-(n) Evolution of the $^{51}$V NMR central transition lines as a function of temperature across different applied pressures. The data primarily capture the spectral evolution below $T_{CDW}$. All the spectra were measured under an external magnetic field of $H = 12$ T.

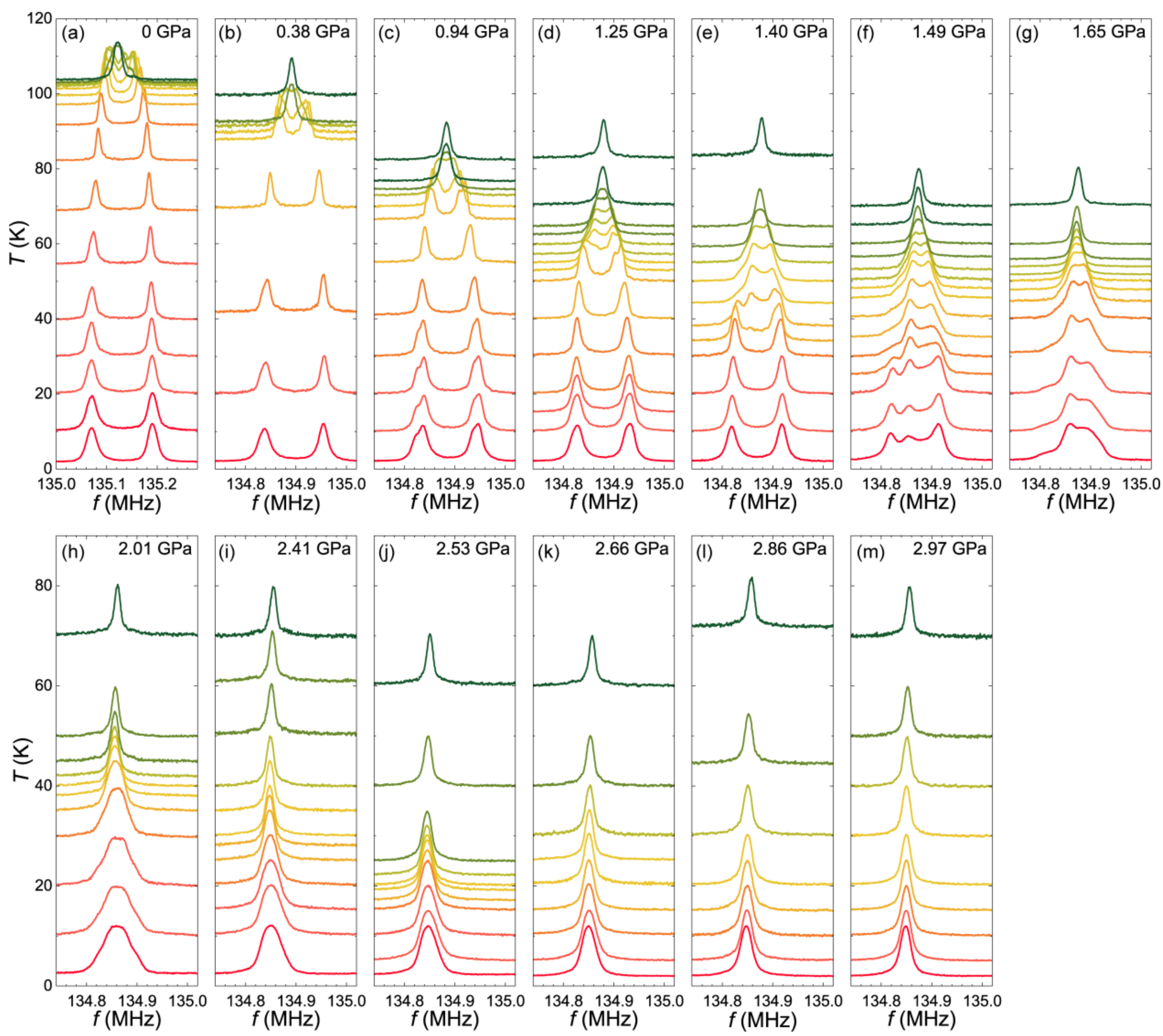

**FIG. S3 *T*-dependent $^{51}$V NMR spectra of $RbV_3Sb_5$ at different pressures.** (a)-(m) Evolution of the $^{51}$V NMR central transition lines as a function of temperature across different applied pressures. The data primarily capture the spectral evolution below $T_{CDW}$. All the spectra were measured under an external magnetic field of $H$ = 12 T.

Section III. Temperature-dependent $^{51}$V NMR $1/T_1T$ under different pressures in $AV_3Sb_5$ (A = K, Rb).

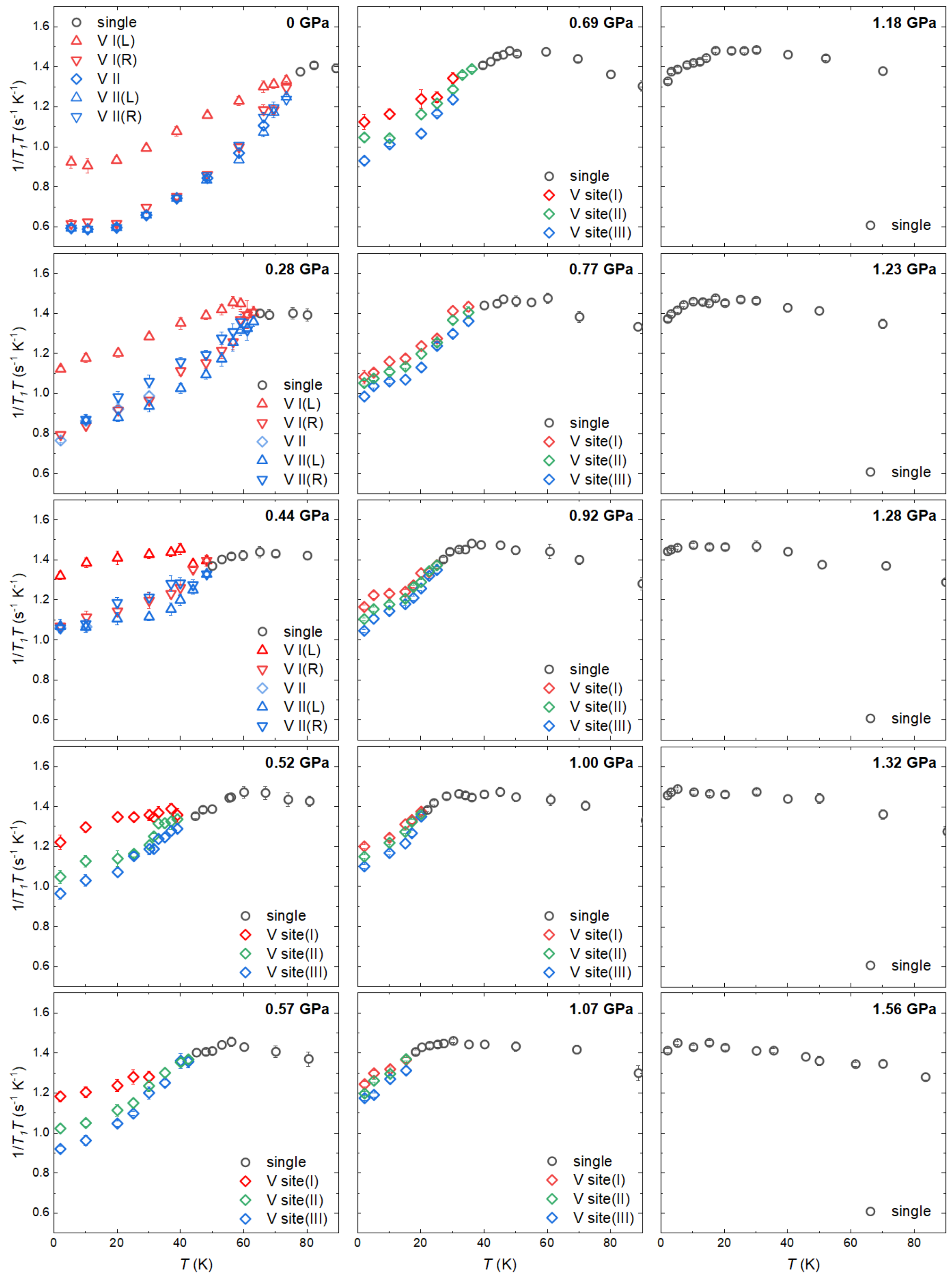

**FIG. S4 *T*-dependent $1/T_1T$ of the $^{51}$V NMR central lines of $KV_3Sb_5$ at different pressures.** The grey open circles represent the $1/T_1T$ extracted from the single-peak central transition lines. Colored open triangles, inverted triangles and diamonds denote the $1/T_1T$ extracted from distinct positions across the broad central transition lines.

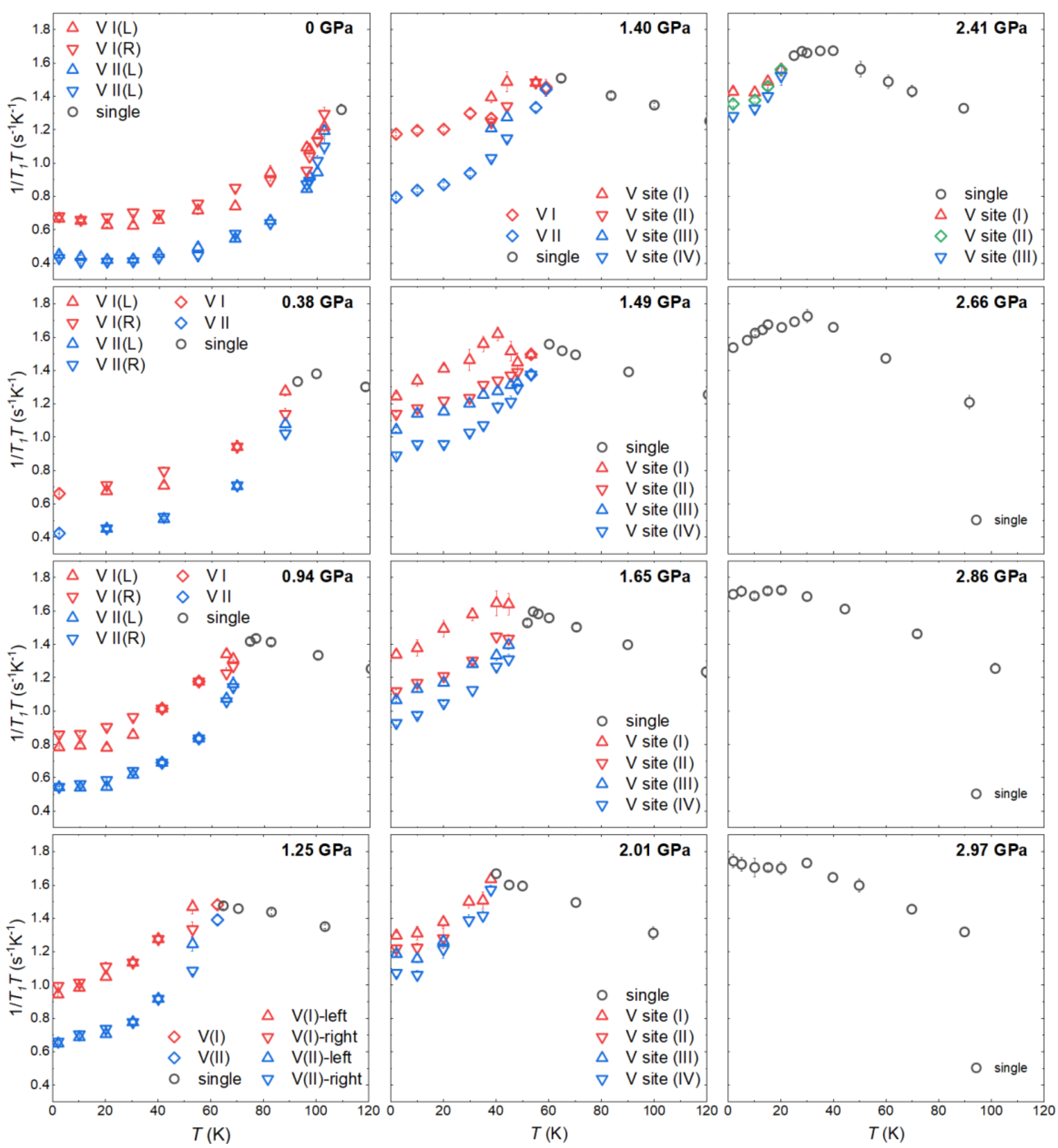

**FIG. S5 *T*-dependent 1/$T_1T$ of the $^{51}$V NMR central lines of $RbV_3Sb_5$ at different pressures.** The grey open circles represent the 1/$T_1T$ extracted from the single-peak central transition lines. Colored open triangles, inverted triangles and diamonds denote the 1/$T_1T$ extracted from distinct positions across the broad central transition lines.

**Section IV. Typical recovery curves for $T_1$ and temperature dependence of parameter $\beta$ in $T_1$ fitting at different pressures in $AV_3Sb_5$ (A = K, Rb).**

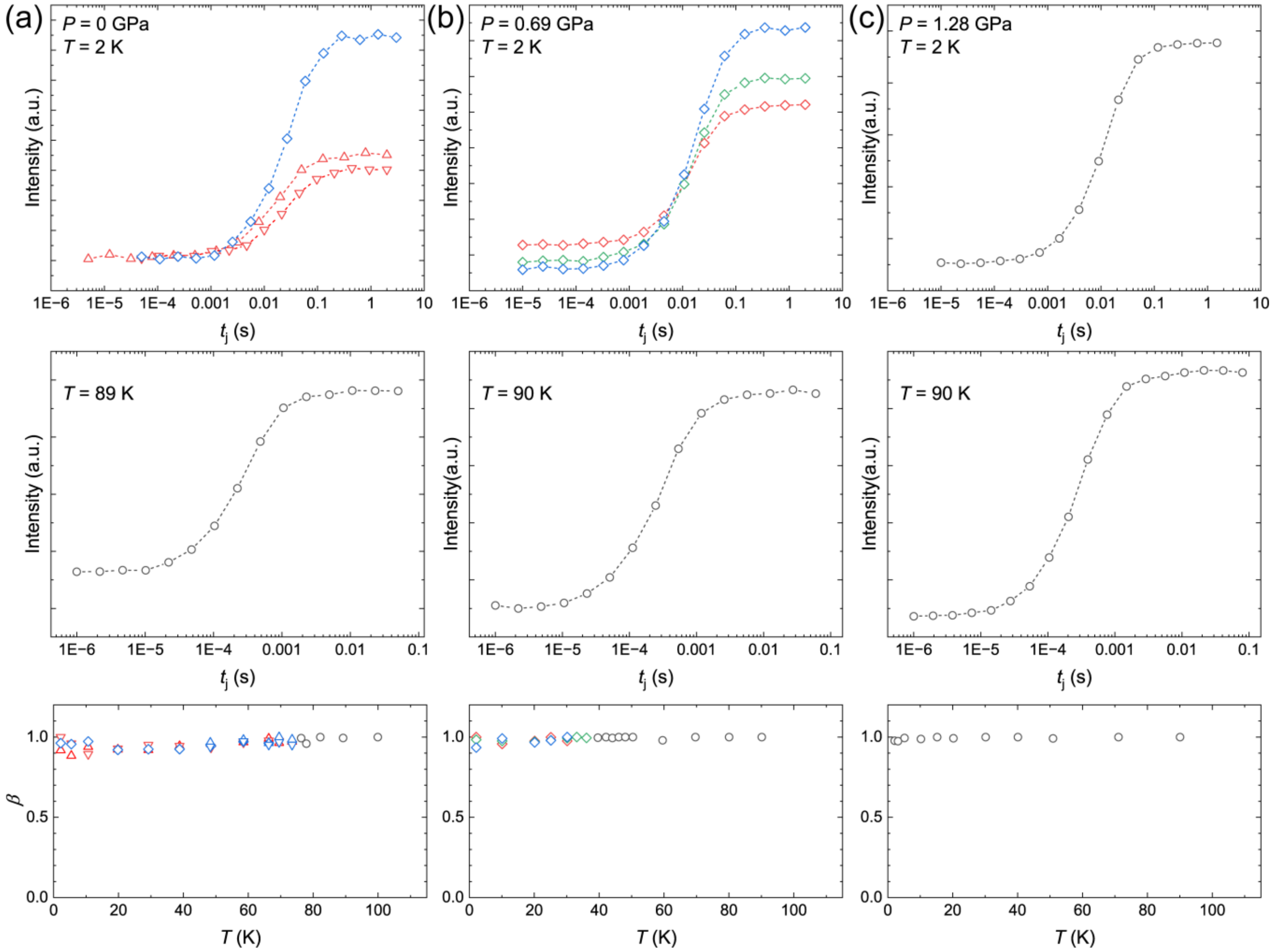

**FIG. S6 Typical $T_1$ recovery curves and $T$-dependent $T_1$ fitting parameter $\beta$ at different pressures in $KV_3Sb_5$.** The $^{51}$V NMR $T_1$ relaxation curves at 2 K (< $T_{CDW}$) and 90 K (> $T_{CDW}$), and the temperature dependence of $\beta$ for $KV_3Sb_5$ at three characteristic pressures: (a) 0 GPa, (b) 0.69 GPa and (c) 1.28 GPa. The colors and shapes of the data points are consistent with those in Fig. S4 at the corresponding pressures.

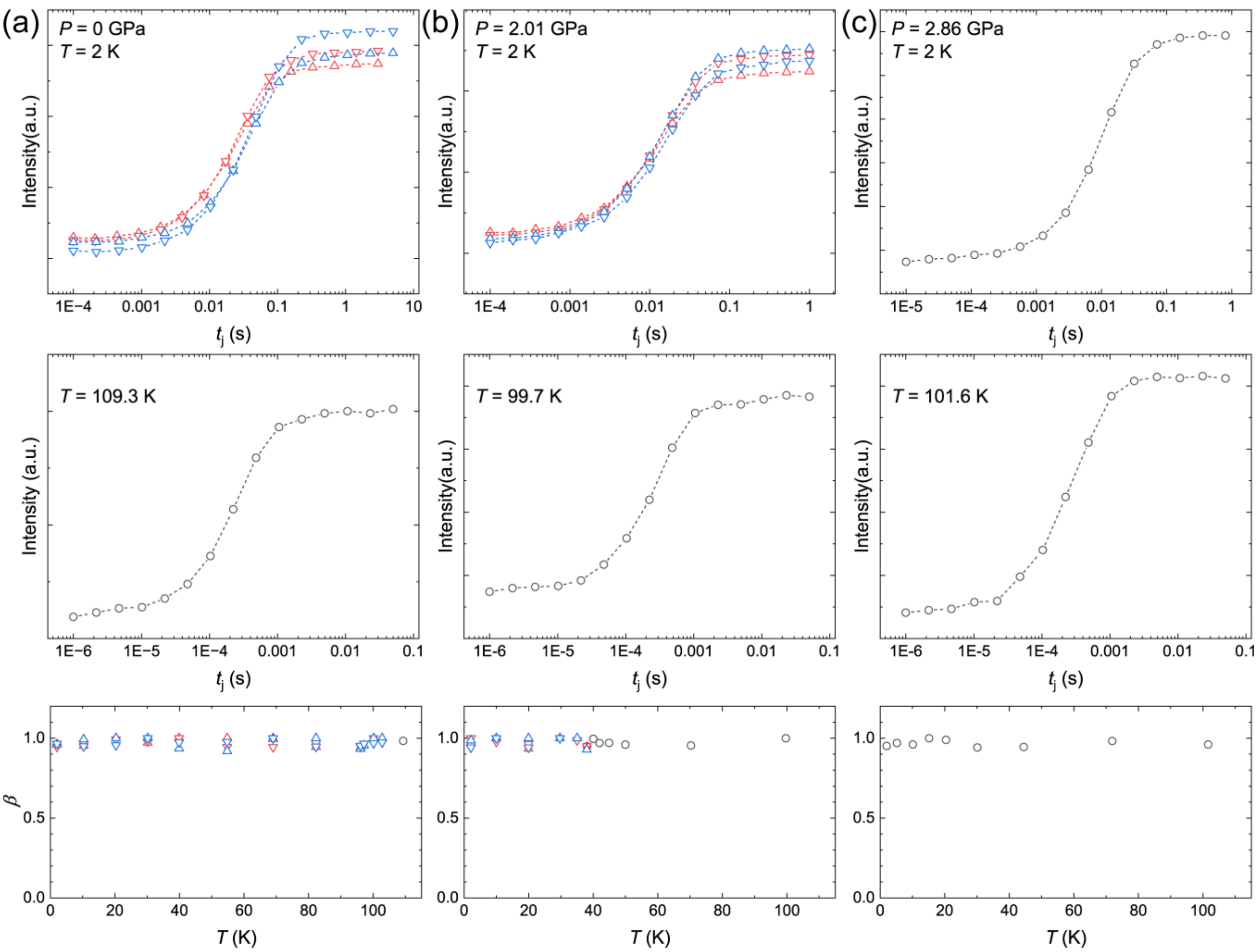

**FIG. S7 Typical $T_1$ recovery curves and $T$-dependent $T_1$ fitting parameter $\beta$ at different pressures in $RbV_3Sb_5$.** The $^{51}$V NMR $T_1$ relaxation curves at 2 K ($< T_{CDW}$) and above $T_{CDW}$, and the temperature dependence of $\beta$ for $RbV_3Sb_5$ at three characteristic pressures: (a) 0 GPa, (b) 2.01 GPa and (c) 2.86 GPa. The colors and shapes of the data points are consistent with those in Fig. S5 at the corresponding pressures.

**Section V. Temperature-dependent $^{51}$V NMR central transition lines in $AV_3Sb_5$ (A = K, Rb and Cs) at $P_{c2}$.**

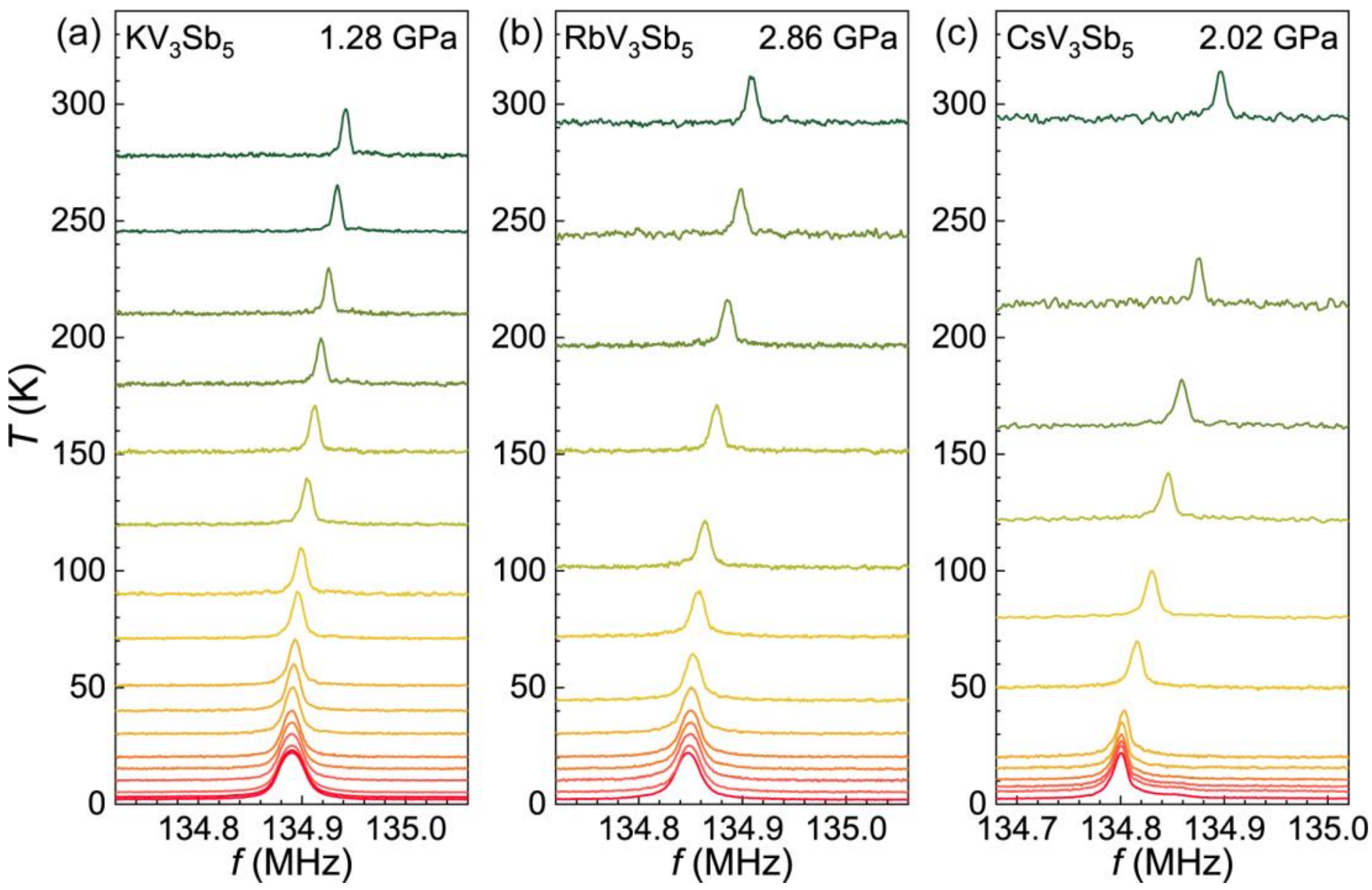

**FIG. S8 *T*-dependent $^{51}$V NMR spectra of (a) $KV_3Sb_5$, (b) $RbV_3Sb_5$ and (c) $CsV_3Sb_5$ at $P_{c2}$.** All the spectra were measured at $H = 12$ T. The Knight shift values are determined from the peak frequencies of the respective central transition lines.